\documentclass[letterpaper,journal]{IEEEtran}

\usepackage{amsthm}
\usepackage[cmex10]{amsmath}
\usepackage{amsfonts}
\usepackage{amssymb}
\usepackage{pmat}
\usepackage{graphicx}
\usepackage{xcolor} 
\usepackage{bbm}
\usepackage{dsfont}
\usepackage{subfigure}
\usepackage{cite}
\usepackage{textcomp}

\def\eps{\varepsilon}
\def\bn{{\boldsymbol{n}}}
\def\bx{{\boldsymbol{x}}}
\def\by{{\boldsymbol{y}}}
\def\bz{{\boldsymbol{z}}}
\def\bC{{\boldsymbol{C}}}
\def\bE{{\boldsymbol{E}}}
\def\bG{{\boldsymbol{G}}}
\def\bH{{\boldsymbol{H}}}
\def\bK{{\boldsymbol{K}}}
\def\bN{{\boldsymbol{N}}}
\def\bQ{{\boldsymbol{Q}}}
\def\bU{{\boldsymbol{U}}}
\def\bV{{\boldsymbol{V}}}
\def\bX{{\boldsymbol{X}}}
\def\calA{{\mathcal{A}}}
\def\calC{{\mathcal{C}}}
\def\calF{{\mathcal{F}}}
\def\calH{{\mathcal{H}}}
\def\calK{{\mathcal{K}}}
\def\calN{{\mathcal{N}}}
\def\calQ{{\mathcal{Q}}}

\newcounter{tempeqncnt}
\newcounter{topeqncnt}

\newtheorem{claim}{Claim}
\newtheorem{corollary}{Corollary}
\theoremstyle{definition} \newtheorem{example}{Example}

\newcommand{\bxa}[1]{\boldsymbol{x}^{(#1)}}

\newcommand{\bxaT}[1]{\boldsymbol{x}^{(#1)\T}}
\newcommand{\bna}[1]{\boldsymbol{n}^{(#1)}}
\newcommand{\bnaT}[1]{\boldsymbol{n}^{(#1)\T}}
\newcommand{\bva}[1]{\boldsymbol{v}^{(#1)}}
\newcommand{\bvaT}[1]{\boldsymbol{v}^{(#1)\T}}
\newcommand{\bvaH}[1]{\boldsymbol{v}^{(#1)\H}}
\newcommand{\bxaPr}[1]{{\boldsymbol{x}'}^{(#1)}}

\newcommand{\bxaPrT}[1]{{\boldsymbol{x}'}^{(#1)\T}}
\newcommand{\bnaPr}[1]{{\boldsymbol{n}'}^{(#1)}}
\newcommand{\bnaPrT}[1]{{\boldsymbol{n}'}^{(#1)\T}}

\newcommand{\s}{\textrm{s}}
\newcommand{\n}{\textrm{n}}
\newcommand{\hs}{h_{\s}}
\newcommand{\hn}{h_{\n}}
\newcommand{\hbars}{\bar{h}_{\s}}
\newcommand{\hbarn}{\bar{h}_{\n}}
\newcommand{\brho}{\boldsymbol{\rho}}
\newcommand{\Ibar}{\bar{I}}
\newcommand{\sd}{\mathrm{SD}}
\newcommand{\Isd}{I_{\sd}}
\newcommand{\Ibarsd}{\Ibar_{\sd}}
\newcommand{\Itil}{\tilde{I}}

\newcommand{\eg}{\textit{e.g.}}

\newcommand{\ie}{\textit{i.e.}}

\newcommand{\perse}{\textit{per se}}
\newcommand{\via}{\textit{via }}

\newcommand{\vide}{\textit{vide }}
\newcommand{\viz}{\textit{viz.}}
\newcommand{\vs}{\textit{vs.}\;}

\newcommand{\norm}[1]{\left\lVert{#1}\right\rVert}
\newcommand{\E}{\textnormal{\textsf{E}}}
\newcommand{\e}{\mathrm{e}}
\renewcommand{\j}{\mathrm{j}}
\newcommand{\T}{\textsf{T}}
\renewcommand{\H}{\textsf{H}}
\newcommand{\tr}{\mathrm{tr}}

\newcommand{\til}[1]{\tilde{#1}}

\renewcommand{\d}{\mathrm{d}}

\newcommand{\complex}[1]{\mathds{C}^{#1}}
\newcommand{\real}[1]{\mathds{R}^{#1}}

\renewcommand{\Re}{\mathrm{Re}}
\renewcommand{\Im}{\mathrm{Im}}

\newcommand{\bzero}[1]{{\boldsymbol{0}}_{#1}}
\newcommand{\bone}[1]{{\boldsymbol{1}}_{#1}}
\newcommand{\boneT}[1]{{\boldsymbol{1}}_{#1}^{\T}}

\newcommand{\brc}[1]{\left( #1 \right)}
\newcommand{\sqbrc}[1]{\left[ #1 \right]}
\newcommand{\figbrc}[1]{\left\{ #1 \right\}}
\newcommand{\bSigma}{\boldsymbol{\Sigma}}

\newcommand{\tilcalQ}{\tilde{\mathcal{Q}}}
\newcommand{\tilrbQ}{\tilde{\boldsymbol{Q}}}

\newcommand{\alp}{\alpha}

\newcommand{\ir}{i_\mathrm{r}}
\newcommand{\ii}{i_\mathrm{i}}
\newcommand{\jr}{j_\mathrm{r}}
\newcommand{\ji}{j_\mathrm{i}}
\newcommand{\lr}{l_\mathrm{r}}
\newcommand{\li}{l_\mathrm{i}}
\newcommand{\br}{b_\mathrm{r}}
\newcommand{\bi}{b_\mathrm{i}}
\newcommand{\ar}{a_\mathrm{r}}
\newcommand{\ai}{a_\mathrm{i}}
\newcommand{\calAr}{\calA_\mathrm{r}}
\newcommand{\calAi}{\calA_\mathrm{i}}

\newcommand{\vw}{{w}}
\newcommand{\vx}{{x}}

\newcommand{\vz}{{z}}

\newcommand{\vbx}{\boldsymbol{x}}
\newcommand{\vby}{\boldsymbol{y}}

\newcommand{\vbG}{\boldsymbol{G}}
\newcommand{\vbH}{\boldsymbol{H}}

\newcommand{\pe}{P_{\mathrm{e}}}

\newcommand{\pebar}{\bar{P}_{\mathrm{e}}}

\newcommand{\bI}{\mathbf{I}}

\newcommand{\MU}{M^{(u)}}
\newcommand{\IU}{I^{(u)}}
\newcommand{\TU}{T^{(u)}}
\newcommand{\pbyxH}{p(\rby|\rbx,\calH)}

\newcommand{\pbx}{p(\rbx)}
\newcommand{\qbx}{q(\rbx')}
\newcommand{\qbyxH}{q(\rby|\rbx',\calH)}

\newcommand{\Ztilu}{\til{Z}^{(u)}}

\newcommand{\etabar}{\eta'}
\newcommand{\epsbar}{\eps'}

\newcommand{\epsk}[1]{\eps_{#1}}

\newcommand{\nuk}[1]{\nu_{#1}}

\newcommand{\rk}[1]{r_{#1}}
\newcommand{\tilrk}[1]{\til{r}_{#1}}
\newcommand{\mk}[1]{m_{#1}}
\newcommand{\tilmk}[1]{\til{m}_{#1}}
\newcommand{\pk}[1]{p_{#1}}
\newcommand{\tilpk}[1]{\til{p}_{#1}}
\newcommand{\qk}[1]{q_{#1}}
\newcommand{\tilqk}[1]{\til{q}_{#1}}

\newcommand{\rbx}{\boldsymbol{x}}
\newcommand{\rby}{\boldsymbol{y}}
\newcommand{\rbz}{\boldsymbol{z}}

\newcommand{\rbC}{\boldsymbol{C}}

\newcommand{\rbX}{\boldsymbol{X}}

\newcommand{\mmse}[1]{\langle #1\rangle}

\newcommand{\Gu}{G^{(u)}}
\newcommand{\muu}{\mu^{(u)}}

\newcommand{\pzeta}{p}

\newcommand{\px}{p}
\newcommand{\qx}{q}

\begin{document}
\IEEEoverridecommandlockouts

\title{Asymptotic Performance Analysis of a $K$-Hop Amplify-and-Forward Relay MIMO Channel}

\author{\IEEEauthorblockN{%
\large{Maksym A.\ Girnyk$^{1}$, Mikko Vehkaper{\"a}$^{2}$, and Lars K.\ Rasmussen$^{3}$} \\}%
\IEEEauthorblockA{%
\normalsize{$^1$Ericsson Research, Stockholm, Sweden\\
$^2$Department of Electronic and Electrical Engineering, University of Sheffield, Sheffield, United Kingdom\\
$^3$School of Electrical Engineering, KTH Royal Institute of Technology, Stockholm, Sweden\\
{\small\begin{tt}maksym.girnyk@ericsson.com\end{tt}, \small\begin{tt}m.vehkapera@sheffield.ac.uk\end{tt},
\small\begin{tt}lars.rasmussen@ieee.org\end{tt}}}}

\thanks{%
The research leading to these results has received funding from the Swedish Research Council (VR). The material of this paper was presented in parts at the IEEE International Symposia on Information Theory in 2013, Istanbul, Turkey, July 2013 and Honolulu, U.S.A., July 2014. The work was done while M.~A.~Girnyk and M.~Vehkaper\"{a} were working at KTH Royal Institute of Technology, Stockholm, Sweden, and Aalto University, Espoo, Finland, respectively.}
}

\maketitle


\begin{abstract} 
The present paper studies the asymptotic performance of multi-hop amplify-and-forward relay multiple-antenna communication channels. Each multi-antenna relay terminal in the considered network amplifies the received signal, sent by a source, and retransmits it upstream towards a destination. Achievable ergodic rates of the relay channel with both jointly optimal detection and decoding and practical separate-decoding receiver architectures for arbitrary signaling schemes, along with average bit error rates for various types of detectors are derived in the regime where the number of antennas at each terminal grows large without a bound. To overcome the difficulty of averaging over channel realizations we apply large-system analysis based on the replica method from statistical physics. The validity of the large-system analysis is further verified through Monte Carlo simulations of realistic finite-sized systems.
\end{abstract}

\begin{IEEEkeywords}
 Relay networks, multi-input multi-output (MIMO), digital modulation, large-system analysis,
 decoupling principle
\end{IEEEkeywords}


\section{Introduction}\label{sec:intro}

\IEEEPARstart{M}{ultiple}-input multiple-output (MIMO) relaying has been proved a promising technology that enables reliable communication with increased coverage and data rates~\cite{telatar1999capacity},~\cite{foschini1998limits}. A wireless sensor network serves as a practically relevant example, where multi-hop (and cluster-based) relaying helps to overcome the low-power-budget constraints~\cite{delcoso2007cooperative}. Thus, understanding of the fundamental limits of multi-hop relay MIMO channels has been regarded as an important milestone in the research efforts within the field of complex cooperative networks~\cite{laneman2004cooperative},~\cite{gesbert2010multi}. In spite of the considerable efforts, however, the capacity of a multi-hop MIMO relay channel, in its most general formulation, remains an open problem.

The relays in a cooperative network may realize different cooperative strategies. A \emph{regenerative} strategy (\eg, decode-and-forward or compress-and-forward~\cite{cover1979capacity},~\cite{kramer2005cooperative}) involves decoding or quantization of the received signal, re-encoding the underlying message and subsequent retransmission upstream. A \emph{non-regenerative} strategy (\eg, amplify-and-forward (AF)~\cite{laneman2004cooperative}) involves direct amplification of the received noisy signal with subsequent retransmission towards the destination. Non-regenerative relaying, being simple in implementation and independent of the modulation schemes at the source terminal, is of particular interest.

The present paper is devoted to a general $K$-hop AF relay MIMO channel in the presence of fast fading, whose achievable rates are characterized by the end-to-end \emph{ergodic mutual information} (MI). Efficient evaluation of the latter is problematic due to computation of the expectation of the MI over the channel realizations, as well as over non-Gaussian input signals in the case the latter are used.

\IEEEpubidadjcol

To overcome these difficulties, several asymptotic approaches, based on the large-system assumption, have been recently proposed. In~\cite{Lee2010}, various asymptotic limits are considered; namely, where the number of antennas grows large either at the source, the destination or the relay terminal, while the number of antennas at the other terminals stays fixed. In other works, the authors apply techniques from random matrix theory to obtain an explicit approximation of the ergodic MI. For instance, in~\cite{Jin2010}, such an explicit expression for the achievable rate of a two-hop AF relay MIMO channel with Gaussian inputs is obtained. Further, assuming that the number of antennas at each terminal within the relay MIMO network grows very large, the authors of~\cite{fawaz2011asymptotic} analyze a multi-hop relay channel with noise only at the destination terminal. Meanwhile, in~\cite{yeh2007asymptotic} the capacity of a general $K$-hop AF relay MIMO channel with Gaussian inputs and noise at every hop is expressed in terms of the limiting eigenvalue distribution of a product of MIMO channel matrices, obtained \via a set of recursive equations. In~\cite[Sec. 3.5]{hoydis2012thesis}, a recursive expression for ergodic MI between the input and output of a $K$-hop AF relay MIMO channel is derived by means of deterministic equivalents introduced in~\cite{hachem2007deterministic}. Another recent paper,~\cite{wen2012efficient}, considers a similar setup while evaluating the average throughput of a multi-hop network under the assumption of full channel state information (CSI) and an SVD-based linear precoder applied at the source terminal.

\begin{figure*}
	\centering
	\includegraphics[width=0.65\linewidth]{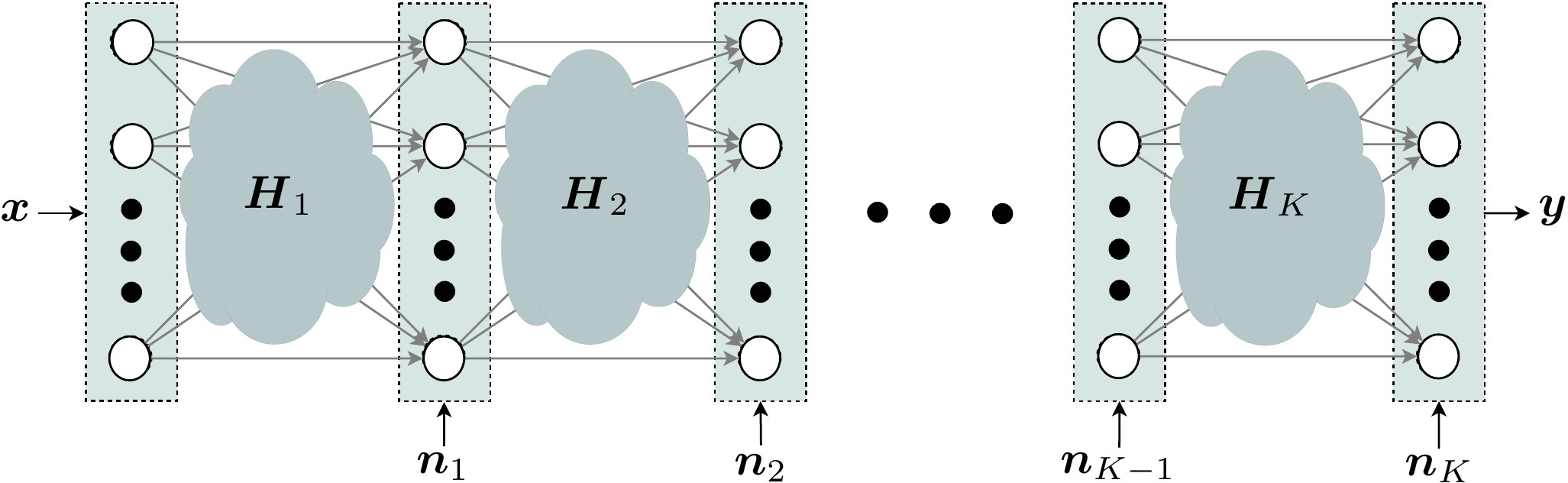}
	\caption{Illustration of a $K$-hop AF relay MIMO channel.} \label{fig:relayKhop}
	\vspace{-0.5cm}
\end{figure*}

Multi-hop AF networks have also been considered in terms of reliability. The latter is usually evaluated in terms of the \emph{average bit error rate} (BER), which, too, yields no closed-form expression for the AF relay MIMO scenario with fading. Thus, in~\cite{girnyk2011myopic}, the instantaneous BER of BPSK transmission is derived in closed-form and the average performance is then obtained through subsequent Monte Carlo estimation. Most of the existing work on closed-form average BER is, however, limited either to one-antenna terminals~\cite{hasna2003end},~\cite{trigui2011closed}, or to bounds on the BER performance~\cite{kim2012performance}. The authors of~\cite{hasna2003end} derive a closed-form expression for average BER of a dual-hop relaying system using various finite-alphabet constellations \via the harmonic mean. An exact closed-form expression for the average BER of single-antenna AF relay system in terms of Lauricella multivariate hypergeometric functions is provided in~\cite{trigui2011closed}. In turn, in~\cite{kim2012performance}, an upper-bound on the average BER of a MIMO AF relay channel is derived using the arithmetic-geometric mean inequality. Furthermore, the large-system BER performance of the linear minimum mean-square error (MMSE) detector in the dual-hop AF relay scenario is analyzed in~\cite{wen2010sum}. Nonetheless, the problem in its general formulation, accounting for other conventional detectors and arbitrary number of hops, $K$, remains open.

The above being said, the present paper provides a framework for efficient performance analysis of a $K$-hop  AF relay MIMO communication channel under general conditions. Namely,
\begin{itemize}
	\item We provide an explicit expression for the end-to-end ergodic MI for a multi-hop channel with arbitrary signaling at the source terminal in the regime, where the numbers of antennas at each terminal grow without bounds at constant ratios. This allows the evaluation of the system spectral efficiency under optimal \emph{joint detection and decoding} (JDD). For Gaussian signals, our results degenerate to those reported in~\cite[Sec. 3.5]{hoydis2012thesis},~\cite{wagner2008large}. For non-Gaussian signals our results partially reduce to those in~\cite{wen2010sum},~\cite{girnyk2013large}.
	\item In addition, we show that the \emph{decoupling principle}, reported by Guo and Verd\'{u}~\cite{guo2005randomly} in the context of a DS-CDMA system, holds also for the multi-hop AF relay MIMO setting. Namely, in the large-system regime, an AF relay MIMO channel, where joint spatial detection at the receiver of the destination terminal is followed by \emph{separate decoding} (SD), decouples into a bank of scalar Gaussian per-stream channels. This allows the characterization of the system performance in the corresponding practically motivated scenario. The obtained results degenerate to those presented in~\cite{wen2010sum},~\cite{wen2007asymptotic},~\cite{girnyk2014asymptotic}.
	\item Furthermore, the framework allows us to determine the average uncoded BER of the system for various types of detection schemes. The corresponding results degenerate to those provided in~\cite{wen2010sum},~\cite{wen2007asymptotic},~\cite{girnyk2014asymptotic},~\cite{muller2003channel}. The results also align with those presented in~\cite{guo2005randomly},~\cite{tanaka2002statistical},~\cite{takeuchi2008asymptotic} for CDMA and MIMO-CDMA systems.
\end{itemize}

Our analysis is based on the replica-symmetric (RS) ansatz of the \emph{replica method} from statistical physics. The method was invented in early 50's by Kac~\cite{kac1968toeplitz}, and it provides a powerful framework for efficient analysis of macroscopic quantities of large many-body systems (\eg, spin glasses~\cite{edwards1975theory},\cite{sherrington1975spinglasses}). Albeit not yet rigorously justified, the method proves efficient in the cases where all other methods fail (\eg, traveling salesman problem~\cite{mezard1986replica}). It was introduced to the field of communications by Tanaka~\cite{tanaka2002statistical}, which inspired lots of subsequent research efforts~\cite{guo2005randomly},~\cite{moustakas2003mimo},~\cite{muller2003channel},~\cite{takeuchi2008asymptotic},~\cite{taricco2008asymptotic}. The method provides a powerful framework for efficient performance analysis of average performance of channels described by a \emph{linear vector model} (such as CDMA or MIMO systems\footnote{In the point-to-point scenario, the result obtained in~\cite{tanaka2002statistical} using the replica method is rigorously proved to be an upper bound to the actual spectral efficiency of a system~\cite{korada2010tight}. Moreover, the obtained therein formula was later partially justified in~\cite{korada2011applications}.}) under general conditions. The method has also gained popularity in other fields of engineering, such as compressed sensing~\cite{kabashima2009typical}, watermarking~\cite{senda2010statistical} and machine learning~\cite{sakata2013time}.

The remainder of the paper is organized as follows. In the following section, we describe the system model and formulate the problem of interest. Next, in Section~\ref{sec:mainResult}, we present the main results of the paper, \viz, an expression for the Helmholtz free energy, along with the decoupling principle, for an equivalent statistical-physics system. In Section~\ref{sec:applications}, we then provide an application of the main result to the AF relay MIMO channel of interest, which enables the evaluation of the performance in terms of the achievable ergodic rate and average BER for various detection schemes. Then, Section~\ref{sec:numericalResults} presents the results of numerical simulations alongside relevant discussion. Finally, in Section~\ref{sec:conclusion}, we draw conclusions. The derivations of the claims are postponed to the appendices.

\emph{Notation:}
Throughout the paper we use upper case bold-faced letters to denote matrices, \eg, $\bX$, with elements denoted by $[\bX]_{i,j}$, lower case bold-faced letters to denote column vectors, \eg, $\bx$, with elements $x_i$, and lower case light-faced letters to denote scalar variables, \eg, $x$. Superscripts $(\cdot)^{\T}$ and $(\cdot)^{\H}$ denote the transpose and Hermitian adjoint operators, respectively. Meanwhile, $\tr\{\bX\}$ and $\det(\bX)$ denote the trace and the determinant of matrix $\bX$. Also, $\bI_M$, $\bzero{M}$ and $\bone{M}$ denote the identity matrix, the all-zeros vector and the all-ones vector of length $M$. Operator $\E\{\cdot\}$ denotes the expectation, $\delta(\cdot)$ denotes the Dirac delta function, while $\otimes$ represents the Kronecker product. Operators $\Re\{\cdot\}$ and $\Im\{\cdot\}$ provide the real and imaginary parts of the argument, respectively. With $\rho$ being the signal-to-noise ratio (SNR), we denote the ergodic MI as $I(\rho)$, the differential entropy of the received signal as $\hs(\rho)$ and the conditional differential entropy as $\hn(\rho)$. The respective quantities with bars on top denote the corresponding asymptotic approximation based on the RS ansatz. Finally, $\Itil_{\eta}(\rho)$ denotes the MI associated with an equivalent decoupled scalar channel with inverse noise variance $\eta$.


\section{System Model}\label{sec:systemModel}

Consider a $K$-hop channel, consisting of $K-1$ multi-antenna relay terminals that assist a multi-antenna source to communicate with a multi-antenna destination. The corresponding setup is depicted in Fig.~\ref{fig:relayKhop}, and it corresponds to that of~\cite{yeh2007asymptotic},~\cite[Sec. 3.5]{hoydis2012thesis}. There is no direct link between the source and the destination, and the terminals operate under a \emph{time-division multiple-access} (TDMA) protocol, so that a single transmitter-receive pair is active at a given time.\footnote{The assumption is motivated by the following reasoning. If all nodes that are not transmitting were to listen for a transmission at all times, backward links in the network would appear. This would lead to an additional problem of scheduling of the transmissions which is outside of the scope of the present paper.} Furthermore, we assume that each relay receives only the signals from the preceding hop. Namely, a symbol sent by the source has to traverse all the $K$ hops before it reaches the destination, where $K$ is a fixed finite number. The source, the destination and the $k$th relay terminal are equipped with $M_0$, $M_K$ and $M_k$ antennas, respectively. AF relaying is employed at each relay, so that the relay simply amplifies and retransmits the received signal upstream without decoding it.

In our flat-fading model, the received signal at the $k$th terminal is given by
\begin{equation}
\label{eqn:channelHopK}
\by_k = \vbH_k \by_{k-1} + \bn_k,
\end{equation}
where $\by_{k-1}$ and $\by_k$ are the input and output of the $k$th channel, respectively. Moreover, $\rho_k$ is the SNR at terminal $k$ and $\beta_{k-1}$ is the normalization constant chosen so that the long-term transmit power constraint at terminal $k-1$ is satisfied, that is,
\begin{equation}\label{eqn:powerConstraint}
\beta_{k-1} \E\figbrc{\by_{k-1}^{\H}\by_{k-1}} \leq M_{k-1}.
\end{equation}
For later convenience, let us define also a set including all the hops $\calK\triangleq\{1,\ldots,K\}$ and group the corresponding SNR values into a vector $\brho \triangleq [\rho_1, \ldots, \rho_K]^{\T}$. The channel matrix between terminals $k-1$ and $k$, $\vbH_k$, is assumed to have circularly symmetric complex Gaussian (CSCG) random entries $[\vbH_k]_{i,j}\sim\mathcal{CN}(0,\rho_k \beta_{k-1} / M_k),\; \forall i,j$, whereas $\bn_k \sim \mathcal{CN}(\mathbf{0}_{M_k},\bI_{M_k})$ is the additive CSCG noise vector at receiver $k$.

The end-to-end input-output relation of the $K$-hop channel can be written as follows:
\begin{equation}
\label{eqn:channelE2E}
\by = \vbG_0^{K-1} \bx + \sum_{k=1}^{K-1} \vbG_k^{K-1} \bn_k + \bn_K,
\end{equation}
where $\bx \triangleq \by_0$ is the input to the relay channel, assumed for the sake of simplicity to have i.i.d. zero-mean unit-variance components distributed according to some probability density function $p(\bx)$, while $\by \triangleq \by_K$, and the corresponding matrices are defined as\footnote{The matrix product notation is defined as $\prod\limits_{k=i}^j \vbH_k \triangleq \vbH_{j} \vbH_{j-1} \cdots \vbH_i$, following~\cite{muller2002asymptotic}, where a similar multi-hop channel was analyzed in terms of the eigenvalue distribution.}
\begin{equation}
\label{eqn:matrixG}
\vbG_i^j \triangleq \prod_{k=i}^j \vbH_{k+1}, \quad j \geq i.
\end{equation}

In this paper, we assume that the destination has full channel state information (CSI),\footnote{CSI acquisition at the destination can, in principle, be done \via conventional pilot-based methods~\cite{tong2003pilot}. Moreover, specific algorithms for pilot design in AF relay MIMO systems, which allow for estimation of the per-hop channels directly at the destination, have recently been proposed~\cite{ma2011pilot},~\cite{kong2011optimal}. The case where the receiver is ignorant of the intermediate channels leads to mismatched decoding \cite{Merhav-etal-1994, Ganti-Lapidoth-Telatar-2000}. For an application of covariance mismatched decoding in the context of large-scale MIMO systems, see for example, \cite{Vehkapera-etal-trcom2015}.} whereas the source and the relays are not aware of the channel states. Hence, the long-term maximum achievable rate is given by the average MI between the input and output of the channel. Define the set of channel matrices $\calH\triangleq\{\vbH_1,\ldots,\vbH_K\}$ and the set of noise realizations $\calN \triangleq \{\bn_1,\ldots,\bn_{K-1}\}$. The end-to-end (normalized) achievable ergodic rate of the channel for a given input distribution $p(\bx)$ is
\begin{equation}
\label{eqn:miErgodic}
I(\brho)\triangleq \frac{1}{M_0}I(\by; \bx) = \hs(\brho) - \hn(\brho),
\end{equation}
where the differential entropy terms are given by\footnote{To account for the TDMA protocol, a factor of $1/K$ is applied to~\eqref{eqn:miErgodic}.}
\begin{subequations}\label{eqn:entropy}
	\begin{align}
		\hs(\brho)\triangleq
		& -\frac{1}{M_0}\E_{\by,\calH} \ln \E_{\bx,\calN} \figbrc{p(\by|\bx,\calH,\calN)}, \label{eqn:entropyY}\\
		\hn(\brho)\triangleq
		& -\frac{1}{M_0}\E_{\by,\bx,\calH} \ln \E_{\calN} \figbrc{p(\by|\bx,\calH,\calN)}, \label{eqn:entropyYGivenX}
	\end{align}
\end{subequations}
with the conditional distribution of the channel being
\begin{equation}
\label{eqn:boltzmannDistribution}
p(\by|\bx,\calH,\calN) = \frac{1}{\pi^{M_K}}\e^{-\norm{\by-\vbG_0^{K-1} \bx - \sum_{k=1}^{K-1} \vbG_k^{K-1} \bn_k}^2}.
\end{equation}

Although~\eqref{eqn:miErgodic} represents the achievable rate of a relay MIMO channel, it assumes optimal joint detection and decoding (JDD), which in practice may be prohibitively complex. A more plausible alternative involves joint spatial detection, followed by a bank of single-user decoders and is referred to as separate decoding (SD) hereafter. In this case, the detector estimates the symbol vector based on the \emph{generalized posterior mean estimator} (GPME)\cite{guo2005randomly}, given by
\begin{equation}
\label{eqn:gpme}
\mmse{\vbx'}_q = \int \rbx' \frac{\qbx \qbyxH}{\int \qbx \qbyxH \d \rbx'} \d \rbx',
\end{equation}
where subscript $q$ reflects the fact that the receiver uses some \emph{postulated} channel law $\qbyxH$ and distribution $\qbx$ for postulated inputs $\vbx'$. The GPME is, in general, suboptimal if the latter two do not match the conditional density $\pbyxH$ and the prior distribution $\pbx$ of the actual channel~\eqref{eqn:channelE2E}, respectively. However, as was shown in~\cite{guo2005randomly}, most of the practically relevant detectors can be regarded as a GPME, optimal for the postulated channel law $\qbyxH$. It was further shown that in order to capture suboptimality of the aforementioned detectors it suffices to postulate a channel with a mismatch only in the noise variance $\sigma^2$ and prior distribution $\qbx$, \ie,
\begin{equation}
\label{eqn:channelPostulated}
\by = \vbG_0^{K-1} \bx' + \sum_{k=1}^{K-1} \vbG_k^{K-1} \bn'_k + \bn'_K,
\end{equation}
where $\vbx'$ is a postulated channel input vector, and $\bn'_k \sim \calC\calN(\bzero{M_k},\sigma^2\bI_{M_k}), \; k\in\calK$ are postulated CSCG noise vectors with variance $\sigma^2$. That is, the conditional density of the postulated channel~\eqref{eqn:channelPostulated} is given by
\begin{equation}
\label{eqn:channelLawPostulated}
q(\by|\bx',\calH,\calN) = \frac{1}{(\pi\sigma^2)^{M_K}}\e^{-\frac{1}{\sigma^2}\norm{\by-\vbG_0^{K-1} \bx' - \sum_{k=1}^{K-1} \vbG_k^{K-1} \bn'_k}^2}.
\end{equation}
The corresponding achievable communication rate of the SD scheme is given by
\begin{equation}
\label{eqn:rateSd}
\Isd(\brho) \triangleq \frac{1}{M_0}I\brc{\mmse{\vbx'};\vbx}.
\end{equation}

\begin{figure*}
	\normalsize
	\setcounter{tempeqncnt}{\value{equation}}
	\vspace{-4pt}
	\setcounter{equation}{15}
	\begin{align} \label{eqn:freeEnergy}
		\calF =&\; \alp_{0,K}\sqbrc{\ln\pi + \frac{1+\eps_K}{\sigma^2+\nu_K} + \ln\brc{\sigma^2 + \nu_K}}
		- \frac{\xi_1}{\eta_1}-\ln\frac{\pi}{\xi_1} - \int p(z;\eta_1) \ln q(z;\xi_1) \d z - \sum_{k=1}^K \alp_{0,k-1}\sqbrc{\xi_{k}\eps_k + \nu_k\frac{\xi_k}{\eta_k}\brc{\xi_k-\eta_k}} \nonumber\\
		&+ \sum_{k=1}^{K-1} \alp_{0,k} \brc{\ln\sqbrc{1+\beta_k\rho_{k+1}\xi_{k+1}\brc{\sigma^2 + \nu_k}} + \beta_k\rho_{k+1} \frac{\xi_{k+1}}{\eta_{k+1}}\frac{\eta_{k+1}\brc{1+\eps_k} - \xi_{k+1}\brc{\sigma^2+\nu_k}}{1+\beta_k\rho_{k+1}\xi_{k+1}\brc{\sigma^2+\nu_k}}}
	\end{align}
	\setcounter{equation}{\value{tempeqncnt}}
	\addtocounter{topeqncnt}{1}
	\hrulefill
	\vspace{-0.5cm}
\end{figure*}

An individually optimal \emph{maximum a posteriori probability} (MAP) detector performs an exhaustive search over all possibly transmitted vectors~\cite{verdu1998multiuser}, yielding the estimate
\begin{equation}
\label{eqn:indOptDetector}
\hat{x}_m = \underset{\chi}{\arg\max} \sum_{\rbx:\;x_m=\chi} p(\by|\bx,\calH).
\end{equation}
Unfortunately, the implementation of such a detector is computationally prohibitive in practice and hence from the practical view-point reduced-complexity alternatives are preferable. The three conventional linear detection schemes considered in this paper are~\cite{joham2005linear}:
\begin{itemize}
	\item \emph{Matched filter} (MF), which maximizes the SNR at the output of the detector, disregarding the interference between the streams.
	\item \emph{Zero forcing} (ZF) filter, which removes the interference, while at the same time enhancing the noise.
	\item \emph{Linear MMSE} (LMMSE) filter, which minimizes the mean-square error (MSE) without constraints on interference, being an optimal linear detector.
	
\end{itemize}

In general, average performance of these detectors over fading channels exhibits no closed-form expression. The analysis is further complicated by the fact that in the AF relay MIMO setting the resulting channel matrix in~\eqref{eqn:channelE2E} becomes a product of the channel matrices of the $K$ hops, yielding non-i.i.d. entries. Note here that in~\cite{muller2002asymptotic} the asymptotic eigenvalue distribution of such a product matrix has been derived, while in~\cite{fawaz2011asymptotic} the ergodic MI has been characterized for a similar large-matrix multi-hop relay channel. However, in both respective models, the noise is added only at the last hop of the relay channel. In contrast, this paper considers a model where the noise is present at each hop, and therefore the two aforementioned results are not applicable. In the next section, we present the developed framework describing the large-system behavior of a $K$-hop AF relay MIMO communication system.


\section{Main Results}\label{sec:mainResult}

Since direct computation of the ergodic MI in~\eqref{eqn:miErgodic} is prohibitive due to the necessity of averaging of the logarithmic terms in~\eqref{eqn:entropy}, some simplifying assumptions have to be invoked to make the problem tractable. In the present paper, we investigate the performance of the system in the \emph{large-system limit} (LSL), meaning that for each intermediate hop $k$ the number of transmit and receive antennas tend to infinity at some constant ratio, \viz, $\forall i,j \; M_{j} = \alpha_{i,j} M_{i} \rightarrow \infty$, where $\alpha_{i,j}$ are finite constants. Moreover, we assume that \emph{self-averaging} property holds, \ie, the randomness of the channel state vanishes in the LSL. This property, being a challenging problem \perse, is yet to be proved in the AF relay  MIMO context.\footnote{For the point-to-point MIMO/CDMA setting with binary inputs both the existence of the LSL and the self-averaging of the normalized free energy were proved in~\cite{korada2010tight}, using the Fekete lemma and the Hamiltonian-perturbation technique.} Hence, the assumption that the property holds is adopted following the existing replica calculus literature~\cite{guo2005randomly},~\cite{muller2003channel},~\cite{takeuchi2008asymptotic}. The above assumptions allow us to state the following two results.

\subsection{Free Energy}
\label{sec:freeEnergy}

The main problem in evaluation of~\eqref{eqn:miErgodic} is due to the expectation operators over the channel states $\calH$ in the differential entropy terms~\eqref{eqn:entropyY} and~\eqref{eqn:entropyYGivenX}. The evaluation becomes particularly difficult in the case of non-Gaussian priors. One way to approach this problem is to use the replica method, which allows us to formulate the upcoming Claim~\ref{thm:freeEnergyAsymptotic}.

As a preliminary step, the differential entropy in~\eqref{eqn:entropyY} is rewritten as follows
\begin{equation}
\label{eqn:entropyFirst}
\hs(\rho) = -\frac{1}{M_0}\E_{\vby,\calH} \ln Z(\vby,\calH),
\end{equation}
where the \emph{partition function} of the corresponding many-body system reads
\begin{equation}
\label{eqn:partitionFunction}
Z(\vby,\calH) \triangleq \E_{\vbx,\calN} \figbrc{\frac{1}{\pi^{M_K}} \e^{-\norm{\by-\vbG_0^{K-1} \bx - \sum_{k=1}^{K-1} \vbG_k^{K-1} \bn_k}^2}},
\end{equation}
with the argument of the exponent being referred to as the \emph{Hamiltonian} of the system. Then, the normalized differential entropy~\eqref{eqn:entropyY} has a meaning of the \emph{Helmholtz free energy} of the system, which can be obtained as~\cite{guo2005randomly}
\begin{align}
	\label{eqn:freeEnergyIdentity}
	\calF =&\; - \frac{1}{M_0} \lim_{u \to 0^+} \frac{\partial}{\partial u} \ln \E_{\vby,\calH} \figbrc{Z^u(\vby,\calH)}.
\end{align}
Note that in the above expression, the expectation has been moved inside the logarithm and it therefore remains to evaluate the $u$th moment of $Z(\vby,\calH)$ for $u \in \real{}$. Considering the LSL, the free energy~\eqref{eqn:freeEnergyIdentity} is evaluated in the claim below.\footnote{The result is obtained using the conventional RS ansatz of the replica method~\cite{tanaka2002statistical},~\cite{guo2005randomly}.}

\begin{claim}
	\label{thm:freeEnergyAsymptotic}
	In the LSL, the free energy~\eqref{eqn:freeEnergyIdentity} is given by~\eqref{eqn:freeEnergy} at the top of the page, where the parameters are obtained from the solution of the following system of equations
	\addtocounter{equation}{1}
	\begin{subequations}
		\label{eqn:fpEqs}
		\begin{align}
			\xi_K =&\; \alp_{K-1,K} (\sigma^2+\nu_K)^{-1},\\
			\eta_K =&\; \alp_{K-1,K} (1+\eps_K)^{-1},\\
			\xi_k =&\; \frac{\alp_{k-1,k} \beta_{k} \rho_{k+1}\xi_{k+1}}{1+\beta_{k}\rho_{k+1}\xi_{k+1}(\sigma^2+\nu_{k})},\\
			\eta_k =&\; \frac{\alp_{k-1,k} \beta_{k} \rho_{k+1}\eta_{k+1}}{1+\beta_{k}\rho_{k+1}\eta_{k+1}(1+\eps_{k})},\\
			\nu_k =&\; \frac{\beta_{k-1}\rho_k(\sigma^2+\nu_{k-1})}{1+\beta_{k-1}\rho_k\xi_k(\sigma^2+\nu_{k-1})},\\
			\eps_k =&\; \frac{\beta_{k-1}\rho_k(1+\eps_{k-1})}{1+\beta_{k-1}\rho_k\eta_k(1+\eps_{k-1})},\\
			\nu_1 =&\; \beta_0 \rho_1 \E_{z,x'} \figbrc{|x' - \mmse{x'}|^2},\label{eqn:msePost}\\
			\eps_1 =&\; \beta_0 \rho_1 \E_{z,x} \figbrc{|x - \mmse{x'}|^2}.\label{eqn:mseActual}
		\end{align}
	\end{subequations}
	Here $\E_{z,x}\figbrc{|x - \mmse{x'}|^2}$ and $\E_{z,x'} \figbrc{|x' - \mmse{x'}|^2}$ denote the MSE and the posterior variance, respectively, associated with the two fixed scalar Gaussian channels given below
	\begin{subequations}
		\label{eqn:chanScalar}
		\begin{align}
			z =&\; \sqrt{\beta_0 \rho_1} x + \frac{w}{\sqrt{\eta_1}},\label{eqn:chanActual}\\
			z =&\; \sqrt{\beta_0 \rho_1} x' + \frac{w'}{\sqrt{\xi_1}},\label{eqn:chanPost}
			\setcounter{topeqncnt}{\value{equation}}
		\end{align}
	\end{subequations}
	where $w,w' \sim \calC\calN (0,1)$. In~\eqref{eqn:mseActual} and~\eqref{eqn:msePost}, $\mmse{x'}$ denotes the MMSE estimate of~\eqref{eqn:chanPost}. In the case of multiple solutions, only that solution minimizing~\eqref{eqn:freeEnergy} is valid.
\end{claim}
\begin{IEEEproof}
	The derivation of the claim is in Appendix~\ref{sec:proofFreeEnergy}.
\end{IEEEproof}

According to~\eqref{eqn:entropyFirst}, the above result allows the characterization of the differential entropy~\eqref{eqn:entropyY} of the source under JDD by setting $\xi_k = \eta_k$ and $\nu_k=\eps_k$ for all $k\in\calK$. Meanwhile, the conditional entropy term \eqref{eqn:entropyYGivenX} is found in a similar way by setting $\eta_1 = \eps_1 = 0$, as will be shown later in Section~\ref{sec:applications}. We also note that a similar result was derived in~\cite{takeuchi2007hierarchical} for a system model that is equivalent to a two-hop channel without noise at the relay. The variables in~\eqref{eqn:fpEqs} describe implicitly similar coupled virtual channels to those explicitly obtained in~\cite{takeuchi2007hierarchical}.

\subsection{Decoupling}
\label{sec:decoupling}

Claim~\ref{thm:freeEnergyAsymptotic} enables the characterization of the performance of an AF relay MIMO system under JDD. To characterize the performance of the SD scheme we formulate the following result.

\begin{claim}
	\label{thm:decoupling}
	Let $\vx_{m}$, $\vx'_m$ and $\mmse{\vx'_m}_q$ denote the $m$th entries of $\vbx$, $\vbx'$ and $\mmse{\vbx'}_q$. In the LSL, the joint distribution of $(\vx_{m},\vx'_m,\mmse{\vx'_m}_q)$ of channels~\eqref{eqn:channelHopK} and~\eqref{eqn:channelPostulated} converges to the joint distribution $(\vx,\vx',\mmse{\vx'}_q)$, associated with channels~\eqref{eqn:chanActual} and~\eqref{eqn:chanPost}.
\end{claim}
\begin{IEEEproof}
	The derivation is presented in Appendix~\ref{sec:proofDecoupling}.
\end{IEEEproof}

The above claim extends the decoupling principle, reported in~\cite{guo2005randomly}, to the $K$-hop AF relay MIMO scenario. It reveals that in the LSL for each data stream $m$ the joint distribution of the input $\vx_{m}$, postulated input $\vx'_{m}$ and output of the GPME $\mmse{\vx'_m}_q$, associated with the original and postulated AF relay MIMO channels~\eqref{eqn:channelHopK} and~\eqref{eqn:channelPostulated}, converges to the joint distribution of the same set of quantities related to the single-user Gaussian scalar channels~\eqref{eqn:chanActual} and~\eqref{eqn:chanPost}. Namely, the channel with the GPME receiver in the LSL decouples into a bank of scalar Gaussian channels fully characterizing its performance.


\section{Performance of a $K$-Hop Relay MIMO Channel}\label{sec:applications}

\subsection{Achievable Rates}
In this section, we apply the results of the asymptotic analysis and evaluate the performance of a $K$-hop MIMO AF relay channel~\eqref{eqn:channelHopK}. Firstly, we evaluate the achievable rate under the assumption of full CSI at the receiver. For this, we evaluate the two differential entropy terms,~\eqref{eqn:entropyY} and~\eqref{eqn:entropyYGivenX}, with help of Claim~\ref{thm:freeEnergyAsymptotic}.

\subsubsection{Joint Detection and Decoding}
If the JDD scheme is employed at the receiver, the postulated input distribution is set exactly the same as the actual distribution, \ie, $q(x) = p(x)$ and $\sigma = 1$. Consequently, we have the following result.
\begin{corollary}
	\label{thm:entropyYAsymptotic}
	In the LSL, the normalized differential entropy~\eqref{eqn:entropyY} is given by
	\begin{align}\label{eqn:entropyYAsymptotic}
		&\hbars(\rho) = \Itil_{\eta_1}(\rho_1) + \alp_{0,K} \ln\brc{1 + \eps_K} - \sum_{k=1}^{K} \alp_{0,k-1} \eta_k \eps_{k} \nonumber\\
		&+ \sum_{k=1}^{K-1} \alp_{0,k} \ln \sqbrc{1 + \rho_{k+1}\beta_{k}\eta_{k+1} (\eps_k + 1)} + \alp_{0,K} (1+\ln\pi).
	\end{align}
	where $\Itil_{\eta_1}(\rho_1)\triangleq I\brc{z; x}$ is the MI between the input $\bx$ and the output $\bz$ of the fixed scalar channel given by
	\begin{align}
		\label{eqn:chanActualNew}
		z =&\; \sqrt{\beta_0 \rho_1} x + \frac{w}{\sqrt{\eta_1}},
	\end{align}
	with $w\sim\calC\calN(0,1)$. The parameters $\eta_k, \eps_k,\; \forall k \in \calK$ satisfy the set of fixed-point equations shown in~\eqref{eqn:fpEqs} with $\xi_k = \eta_k$ and $\nu_k=\eps_k$ for all $k\in\calK$, where the term $\eps_1= \beta_0 \rho_1 \E_{z,x} \figbrc{|x - \mmse{x}|^2}$ reflects the MMSE of the fixed scalar channel~\eqref{eqn:chanActualNew}. Furthermore, the normalized conditional entropy~\eqref{eqn:entropyYGivenX} in the LSL may be computed directly by~\eqref{eqn:entropyYAsymptotic} and~\eqref{eqn:fpEqs} by setting $\eta_1 = \eps_1 = 0$ beforehand. In the case of multiple solutions, the one minimizing the differential entropy of interest is valid.
\end{corollary}

\begin{IEEEproof}
	The result follows from Claim~\ref{thm:freeEnergyAsymptotic} for the special case of $q(x) = p(x)$ and $\sigma = 1$. Due to the RS assumption we set $\eta_k = \xi_k$ and $\eps_k = \nu_k$ for all $k\in\calK$.\footnote{It is noteworthy that $p(x)=q(x)$ and $\sigma=1$ do not necessarily imply $\eta_{k}=\xi_{k}$ and $\eps_{k}=\nu_{k}$. The latter is based on the assumption that RS holds for the JDD scheme, which has been rigorously justified for the CDMA systems with BPSK inputs in~\cite{nishimori2001comment}. In the unlikely case where the assumption does not hold, there may exist other solutions for which $\eta_{k}\neq\xi_{k}$ and $\eps_{k}\neq\nu_{k}$.} Then, in the conditional entropy term~\eqref{eqn:entropyYGivenX}, the inner expectation is taken only over $\calN$. Hence one has to proceed in exactly the same way as before, but without replication of $\bx$ and $\bx'$, as done in~\eqref{eqn:replication} in Appendix~\ref{sec:proofFreeEnergy}. When the free energy is derived, one sets $\eta_k = \xi_k$ and $\eps_k = \nu_k$ for all $k\in\calK$ and $\eta_1 = \eps_1 = 0$ to obtain the asymptotic approximation for~\eqref{eqn:entropyYGivenX}.
\end{IEEEproof}

The expectation in~\eqref{eqn:mseActual} is taken over the joint distribution $p(z,x;\eta_1)$, and hence $\eps_1$ can be seen as the MMSE of the fixed scalar channel~\eqref{eqn:chanActual}. The entropy term $\hbars(\brho)$, given in~\eqref{eqn:entropyYAsymptotic}, represents the amount of information contributed by the transmitted signal, $\bx$, and by the noise components, $\bn_k$, added at each hop. Note that both the MMSE and MI terms are relatively easy to compute since they are associated with fixed scalar channels~\eqref{eqn:chanActual} and~\eqref{eqn:chanPost}. Meanwhile, the differential entropy $\hbarn(\brho)$ represents the amount of information discarded at the destination terminal due to noise removal, and hence does not contain terms related to the signal vector $\vbx$.

\subsubsection{Joint Detection and Separate Decoding}
Based on the decoupling result of Claim~\ref{thm:decoupling}, along with the fact that for the SD scheme statistical properties of the MIMO channel~\eqref{eqn:channelHopK} are completely characterized by the conditional joint distribution $p(\vx_{m},\vx'_m,\mmse{\vx'_m}_q|\calH)$, we have the following result.

\begin{corollary}
	\label{thm:miSd}
	In the LSL, the single-user achievable rate under SD converges to
	\begin{align}\label{eqn:miSd}
		\Ibarsd(\brho)  = -\int \pzeta(z;\eta_1) \ln \pzeta(z;\eta_1) \d z -\ln\frac{\pi\e}{\eta_1}.
	\end{align}
\end{corollary}

Knowing both the differential entropy terms in Claim~\ref{thm:entropyYAsymptotic} and MI~\eqref{eqn:miSd} above, one can directly compute the achievable rate as a function of SNRs $\brho$ for both the JDD and SD schemes. For instance, the following examples present the expressions for the MMSE and MI for three particularly relevant signal constellations.

\begin{example}[\emph{Gaussian inputs}]
	When $p(x)$ is the standard complex Gaussian density, the MMSE term~\eqref{eqn:mseActual} is given by
	\begin{equation}
	\eps_1 = \frac{\rho_1 \beta_0}{1 + \rho_1\beta_0\eta_1},
	\end{equation}
	and the MI between the input and output of~\eqref{eqn:chanActual} reads
	\begin{equation}
	\Itil_{\eta_1}(\rho_1) = \ln \brc{1 + \rho_1\beta_0\eta_1}.
	\end{equation}
\end{example}

\begin{example}[\emph{QPSK inputs}]
	For the QPSK constellation we have $p(x) = 1/4$ for all $x\in\{\pm\frac{1}{\sqrt{2}} \pm \frac{\j}{\sqrt{2}}\}$. The MMSE term~\eqref{eqn:mseActual} reads
	\begin{equation}
	\label{eqn:epsQpsk}
	\eps_1 = \rho_1\beta_0 - \frac{\rho_1\beta_0}{\sqrt{2\pi}}\int_{\mathds{R}} \e^{-\frac{z^2}{2}}\tanh\brc{\beta_0\rho_1\eta_1 + \sqrt{\beta_0\rho_1\eta_1}z}\d z,
	\end{equation}
	and by the I-MMSE relation~\cite{guo2005mutual}, the MI between the output and the input of~\eqref{eqn:chanActual} is evaluated as
	\begin{align}
		\Itil_{\eta_1}&(\rho_1) = 2 \rho_1\beta_0\nonumber\\
		& - \frac{2}{\sqrt{2\pi}}\int_{\mathds{R}} \e^{-\frac{z^2}{2}} \ln \cosh(\beta_0\rho_1\eta_1 + \sqrt{\beta_0\rho_1\eta_1}z)\d z.
	\end{align}
\end{example}

Finally, comparing the results obtained for JDD and SD, we are able to quantify the loss due to separation of decoding.
\setcounter{equation}{23}
\begin{corollary}
	\label{thm:lossSdd}
	In the LSL, the information loss due to separation of detection and decoding is given by
	\begin{align}
		\Ibar(\brho) &- \Ibarsd(\brho) = \sum_{k=1}^{K-1} \alp_{0,k} \ln \sqbrc{\frac{1 + \rho_{k+1}\beta_{k}\eta_{k+1} (\eps_k + 1)}{1 + \rho_{k+1}\beta_{k}\etabar_{k+1} (\epsbar_k + 1)}} \nonumber\\
		& +\alp_{0,K} \ln\sqbrc{\frac{1 + \eps_K}{1 + \epsbar_K}}\! -\! \sum_{k=1}^{K} \alp_{0,k-1} \brc{\eta_k \eps_{k} - \etabar_k \epsbar_{k} } ,
	\end{align}
	where $\eta_k$ and $\eps_k$ correspond to the entropy term~\eqref{eqn:entropyY}, while $\etabar_k$ and $\epsbar_k$ correspond to~\eqref{eqn:entropyYGivenX}; both obtained according to Claim~\ref{thm:entropyYAsymptotic}.
\end{corollary}

\subsection{Bit Error Rate}
Since by Claim~\ref{thm:decoupling} the performance of the given system with SD is fully described by the equivalent scalar channels~\eqref{eqn:chanScalar}, we can evaluate the average BER of uncoded transmission. The following result summarizes the finding for a QPSK constellation.
\begin{corollary}
	In the LSL, the average BER of the AF relay MIMO system operating with QPSK inputs is given as
	\begin{align}\label{eqn:berAsymptotic}
		\pebar(\brho) = Q\brc{\sqrt{\beta_0\rho_1 \eta_1}},
	\end{align}
	where $\eta_1$ is a solution to the fixed-point equation system~\eqref{eqn:fpEqs}.
\end{corollary}

\begin{example}[\emph{Individually optimal detector}]
	To realize MAP detection, the receiver needs to postulate the actual distribution, \ie, $\qx(x) = \px(x)$ and $\sigma = 1$. This leads to $\xi_k = \eta_k$ and $\nu_k = \eps_k$ for $k \in\calK$. Moreover, $\eps_1$ in~\eqref{eqn:mseActual} depends on the input constellation and has to be evaluated numerically. For instance, for QPSK signals, the parameter is given by~\eqref{eqn:epsQpsk}. In case of multiple solutions, the one minimizing the free energy~\eqref{eqn:freeEnergy} should be chosen.
\end{example}

\begin{example}[\emph{Linear detectors}]
	For a linear detector, the postulated input distribution $\qx(x)$ is set to be standard Gaussian, which yields the MSE of~\eqref{eqn:chanActual} and posterior variance of~\eqref{eqn:chanPost}
	\begin{subequations}\label{eqn:epsNu}
		\begin{align}
			\eps_1 =&\; \frac{\beta_0\rho_1(\eta_1+\beta_0\rho_1\xi_1^2)}{\eta_1(1+\beta_0\rho_1\xi_1)^2},\\
			\nu_1 =&\; \beta_0\rho_1(1+\beta_0\rho_1\xi_1)^{-1}.
		\end{align}
	\end{subequations}
	Then, by tuning the parameter $\sigma$, we can obtain the three popular linear detectors:
	\begin{itemize}
		\item When $\sigma \to \infty$, the output of GPME converges to the output of the MF.
		\item When $\sigma \to 0$, the output of GPME tends to that of the ZF detector.
		\item When $\sigma \to 1$, the output of GPME provides the output of the LMMSE detector.
	\end{itemize}
	Finally, to obtain an approximation for the average BER, one has to plug the corresponding value of $\sigma$ into~\eqref{eqn:fpEqs} and find the unique solution for the system of fixed-point equations. Then, the resulting $\eta_1$ is plugged into~\eqref{eqn:berAsymptotic} and the average BER is acquired.
\end{example}


\section{Numerical Results}\label{sec:numericalResults}

\subsection{Achievable Rates}

To support the main result, we simulate a network with $K=3$ hops. To satisfy the power constraint~\eqref{eqn:powerConstraint}, we set the normalization coefficients $\beta_k$, according to
\begin{equation}
\beta_k = \brc{1+\rho_k}^{-1}, \quad \forall k \in \calK.
\end{equation}

\subsubsection{Joint detection and decoding}
\begin{figure}
	\centering
	\includegraphics[height=7cm,width=8cm]{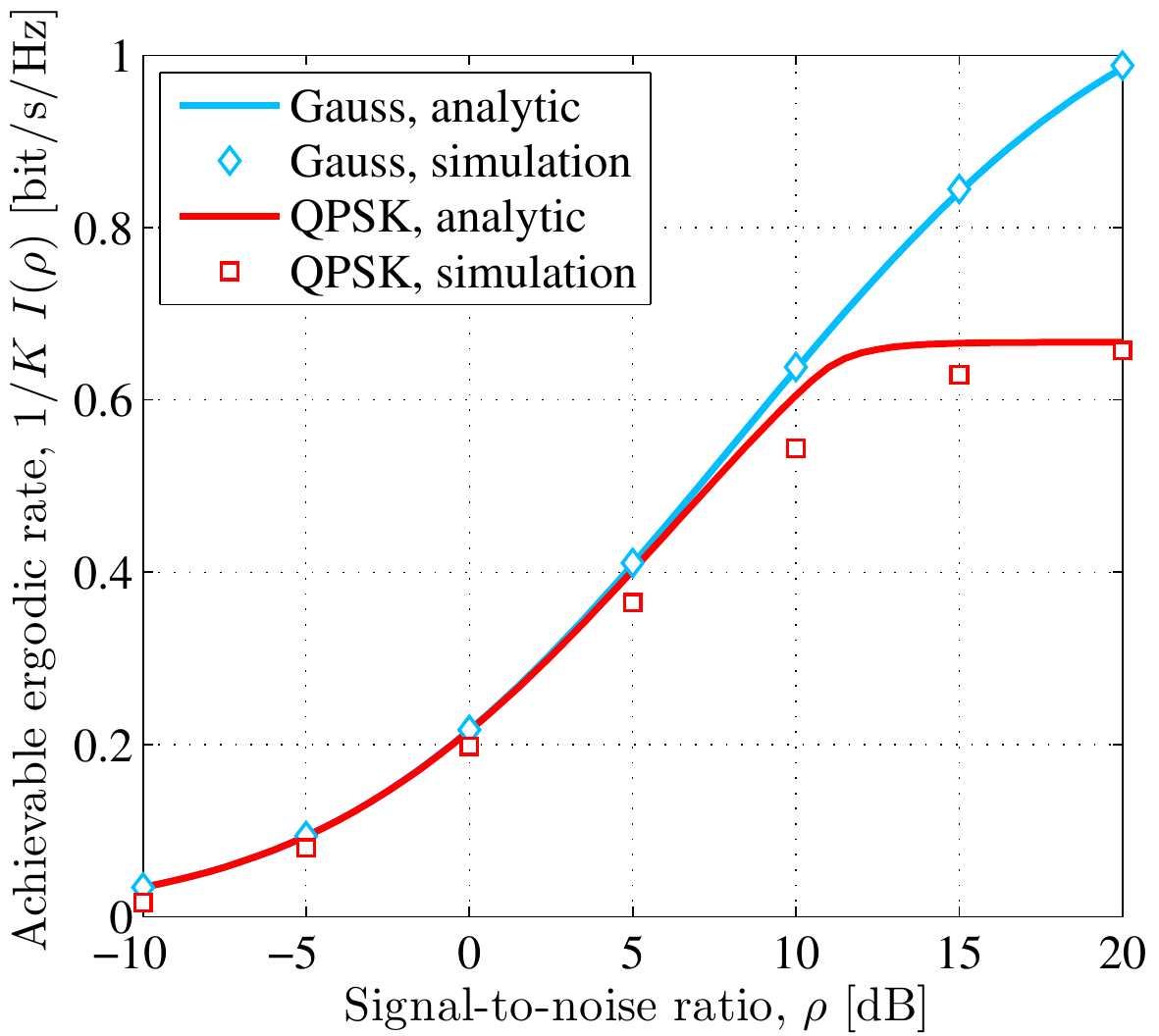}
	\caption{Per-dimension achievable rate \vs SNR for an AF relay channel with $K=3$ hops. A factor of $1/K$ in front of the rate is due to the TDMA protocol. Transmit SNR for the first hop is given by $\rho_1 = \rho$, while the rest of the SNRs are set as $\rho_k = 20$ dB for $k\in\{2,3\}$. Terminals are equipped with $M_k = 8$ antennas, for $k\in\{0,\ldots,3\}$. Solid curves denote the analytic results, while markers denote the results of Monte Carlo simulation.}
	\vspace{-0.5cm}
	\label{fig:resultMi3HopQpskJddMcSnr}
\end{figure}

\begin{figure}
	\centering
	\includegraphics[height=7cm,width=8cm]{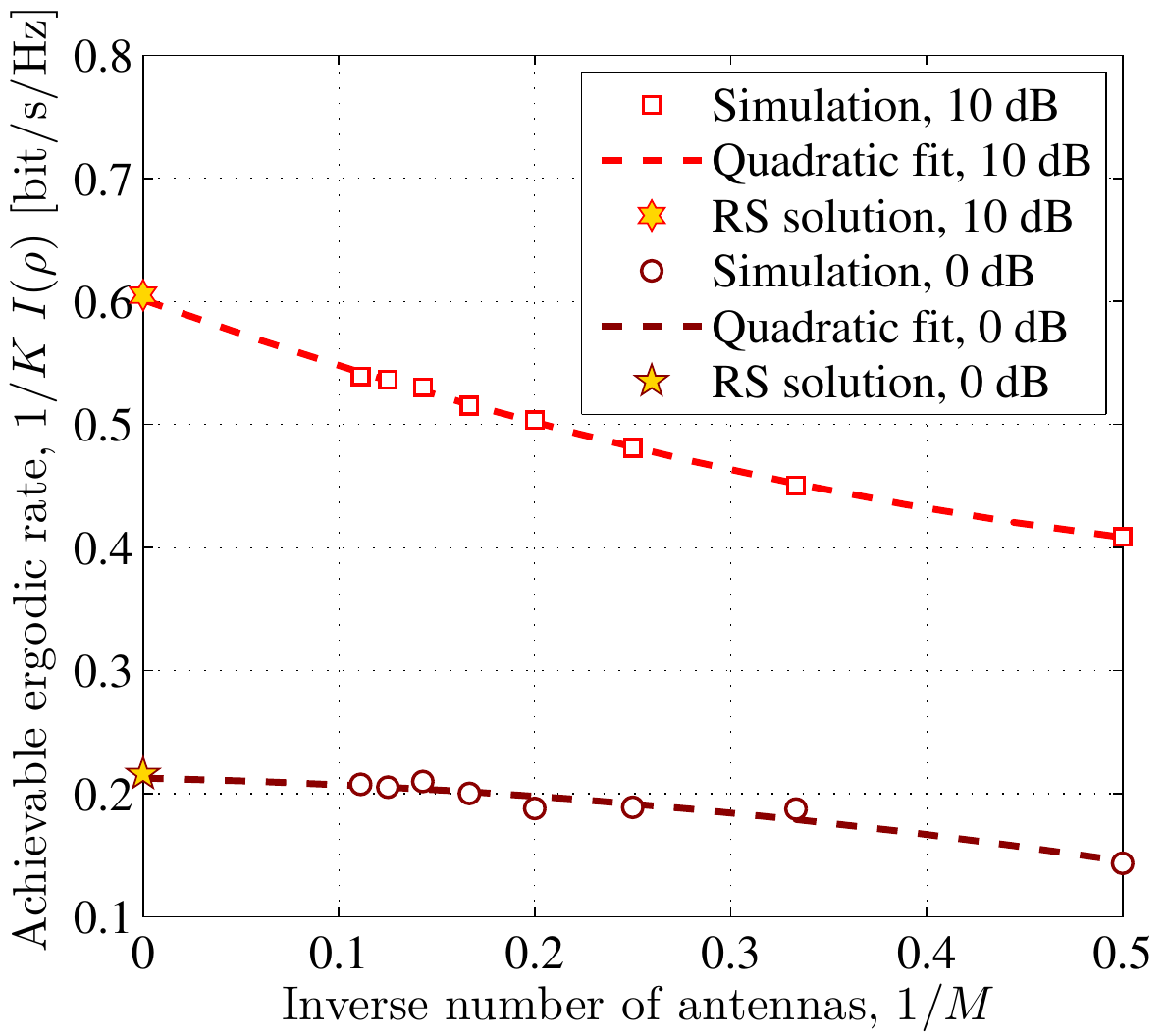}
	\caption{Per-dimension achievable rate achieved by QPSK signaling \vs inverse number of antennas at terminals for an AF relay channel with $K=3$ hops. A factor of $1/K$ in front of the rate is due to the TDMA protocol. Transmit SNR for the first hop is given by $\rho_1 \in \{0,10\}$ dB, while the rest of the SNRs are set as $\rho_k = 20$ dB for $k\in\{2,3\}$. Terminals are equipped with $M_k = M$ antennas, for $k\in\{0,\ldots,3\}$ and $M\in\{2,\ldots,9\}$. The markers denote the results of Monte Carlo simulation, while the dashed lines denote the results of quadratic curve fitting to these points. The star markers at $1/M \to 0$ denote the predictions obtained by the replica analysis in the LSL.}
	\vspace{-0.5cm}
	\label{fig:resultMi3HopQpskJddMc1M}
\end{figure}

For the first simulations setup let us fix $M_0 = M_1 = M_2 = M_3 = 8$, so that each terminal has eight antennas, and let SNRs be $\rho_2 = \rho_3 = 20$ dB, while varying the SNR of the first hop $\rho_1$. In Fig.~\ref{fig:resultMi3HopQpskJddMcSnr}, we plot the ergodic rates achievable with JDD at the receiver for two types of channel inputs: Gaussian and QPSK signals. The rate loss of $1/K$ here is due to the TDMA protocol. Solid lines reflect the asymptotic results, while markers represent the numerical averaging \via Monte Carlo simulations over at least 1000 channel realizations.\footnote{Discrete constellations usually require a much higher number of Monte Carlo iterations to obtain stable results. For instance $2.5\cdot 10^{10}$ iterations have been simulated to produce each point in Fig.~\ref{fig:resultMi3HopQpskJddMcSnr} and Fig.~\ref{fig:resultMi3HopQpskJddMc1M}} We note that for the case of Gaussian inputs the approximation matches the simulations perfectly. Moreover, quite expectedly, it reproduces the result obtained in~\cite[Sec. 3.5.3]{hoydis2012thesis}. Meanwhile, for the case of QPSK inputs there is a slight gap in the middle- and high-SNR region, which decreases with increasing the actual number of antennas at each terminal.

To illustrate the latter argument, we plot in Fig.~\ref{fig:resultMi3HopQpskJddMc1M} the simulated values of the ergodic rates achievable with QPSK inputs and a JDD receiver \vs $1/M$ for $M \in \{2,\ldots, 9\}$, where $M = M_0 = M_1 = M_2 = M_3$, so that each terminal has $M$ antennas. The SNRs are chosen as $\rho_1 \in \{0, 10\}$ dB and $\rho_2 = \rho_3 = 20$ dB. The values of $\rho_1$ are chosen to illustrate the convergence in the cases where the numerical results are close and far w.r.t. the asymptotic curve in Fig.~\ref{fig:resultMi3HopQpskJddMcSnr}. The star markers at $1/M \to 0$ represent the asymptotic results obtained using Corollary 1 in the LSL and quadratic curves are fitted to the simulated data using non-linear least-squares regression. From the extrapolation one can see that in both cases the simulated ergodic rate approaches the RS solution as $M \to \infty$.

\begin{figure}
	\centering
	\includegraphics[height=7cm,width=8cm]{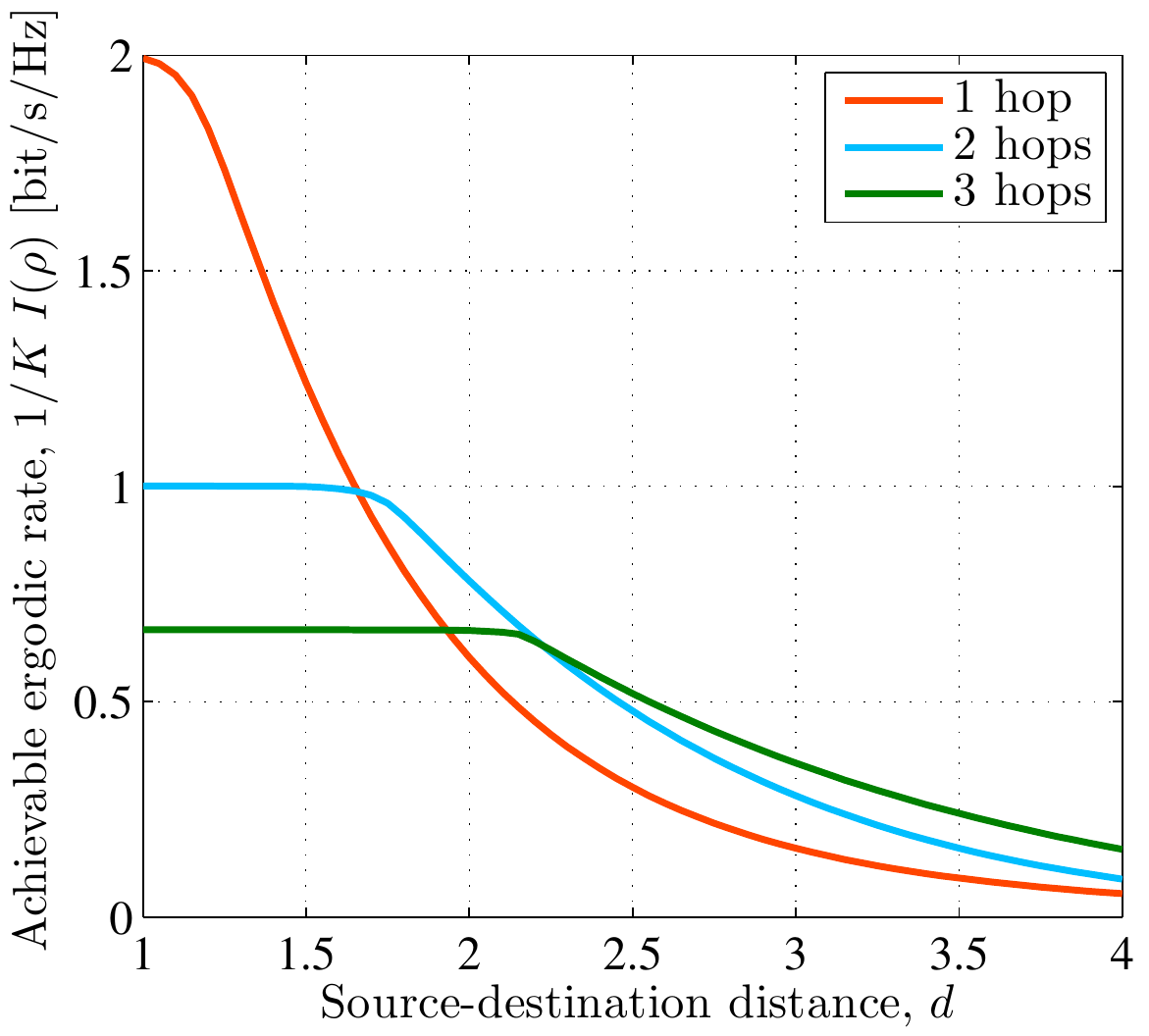}
	\caption{Per-dimension achievable rate under QPSK signaling \vs the distance between the source and destination terminals for a $K$-hop AF relay MIMO channel with $K\in\{1,2,3\}$. A factor of $1/K$ in front of the rate is due to the TDMA protocol. Pathloss $\gamma_k = d_k^{-\alpha}$ is incorporated into SNRs $\rho_k = 10$ dB, for $k\in\{1,\ldots,K\}$. Solid curves denote analytic results, while the dashed curve denotes the best strategy.}
	\vspace{-0.5cm}
	\label{fig:resultMi3HopQpskDist}
\end{figure}

Next, to illustrate the usefulness of the result we incorporate pathloss $\gamma_k = d_k^{-\alpha}$ into the SNRs $\rho_k$ (set to 10 dB for all $k\in\calK$) where $d_k$ is the distance between terminals $k-1$ and $k$, and $\alpha = 4$ is the pathloss exponent. In Fig.~\ref{fig:resultMi3HopQpskDist}, we plot the achievable rate of JDD as a function of the distance, $d$, between the source and destination terminals. The three curves correspond to $K = 1,2,3$. The relays are added in such a way that all the terminals of the network are equidistant. Notably, for different values of $d$, different numbers of hops provide higher achievable ergodic rate (depicted by a dashed line). Thus, one could, in principle, use this information to select the most suitable number of relays or their positions. This selection, however, falls outside the scope of the present paper.

\subsubsection{Separate decoding}
\begin{figure}
	\centering
	\includegraphics[height=7cm,width=8cm]{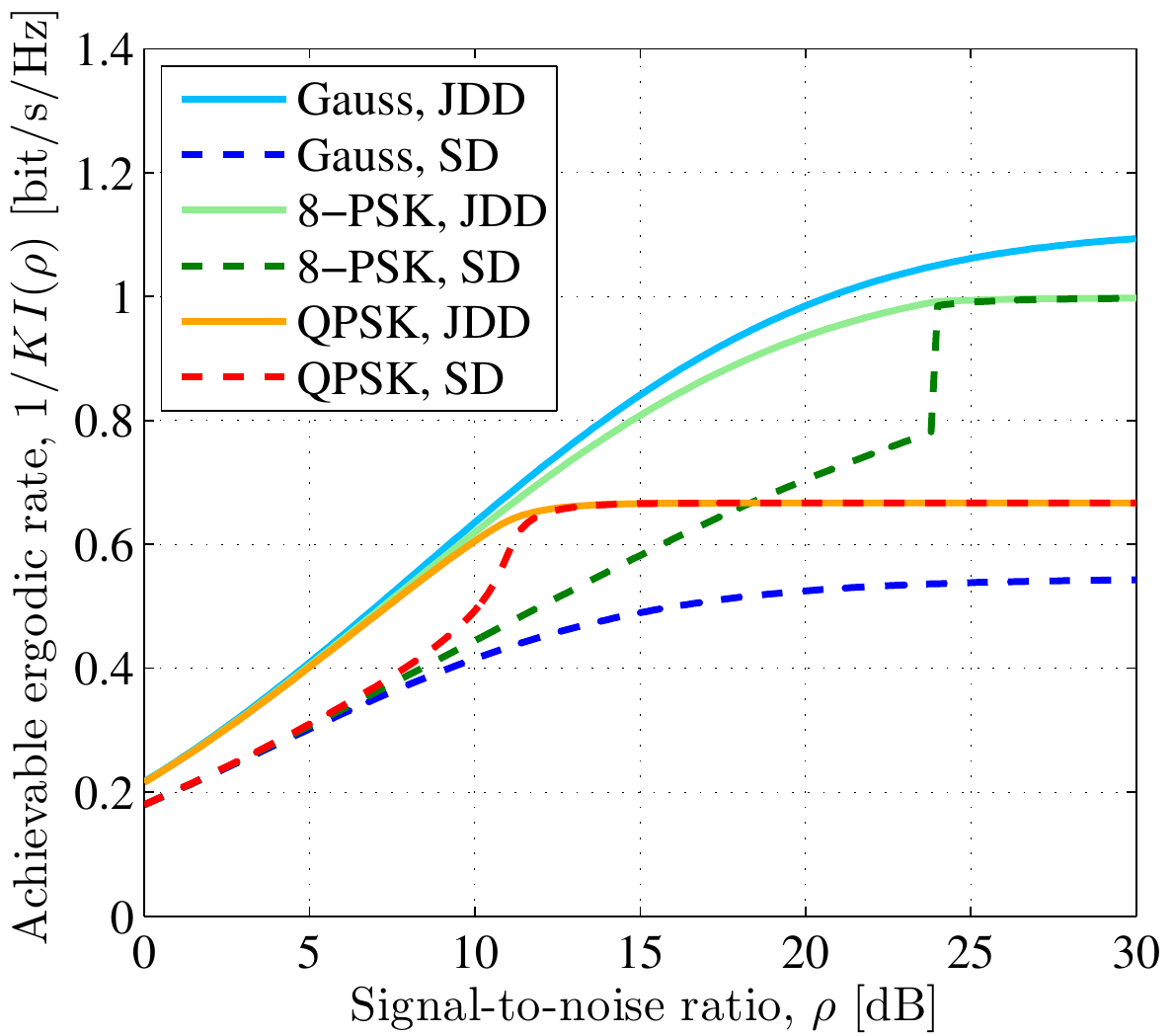}
	\caption{Per-dimension achievable rate \vs SNR for an AF relay channel with $K=3$ hops. A factor of $1/K$ in front of the rate is due to the TDMA protocol. The SNR of the first hop is given by $\rho_1 = \rho$, while the rest of the SNRs are set as $\rho_k = 20$ dB for $k\in\{2,3\}$. Terminals are equipped with $M_k = 8$ antennas, for all $k\in\{0,\ldots,3\}$. Solid curves denote performance of the JDD scheme, while dashed lines denote that of SD.}
	\vspace{-0.5cm}
	\label{fig:resultMi3Hop8PskSdSnr}
\end{figure}

In Fig.~\ref{fig:resultMi3Hop8PskSdSnr}, for the same antenna setup we plot achievable rates of both JDD and SD schemes as functions of SNR for different signaling schemes (Gaussian, QPSK and 8-PSK). We observe that at low SNR the performance curves of discrete constellations tend to follow the respective curves (JDD and SD), related to Gaussian signaling schemes. Then at certain SNR values, the performance curves, corresponding to the discrete inputs, suddenly switch to the entropy-limited regime. Such a sudden change in performance indicates the occurrence of a \emph{phase transition} at a certain SNR threshold. A physical analogy to such a behavior is freezing water or the hysteresis of a ferro-magnetic material~\cite{bertotti1998hysteresis}. From a practical point of view, this can be explained by the sparsity of a discrete constellation, which helps to identify symbols perfectly once the SNR is sufficiently high. Similar behavior was also observed in the context of MIMO~\cite{muller2003channel} and CDMA systems~\cite{tanaka2002statistical}, as well as iterative turbo coding~\cite{montanari2000turbo}. Notably, for the SD scheme with antenna ratios $\alp_{0,1},\alp_{1,2} > 1$, Gaussian signaling is no longer optimal, and it is outperformed by discrete constellations, as previously observed in~\cite{guo2005randomly},~\cite{muller2004capacity}.

\begin{figure}[t]
	\begin{center}
		\subfigure[Gaussian inputs]{\includegraphics[height=7cm,width=8cm]{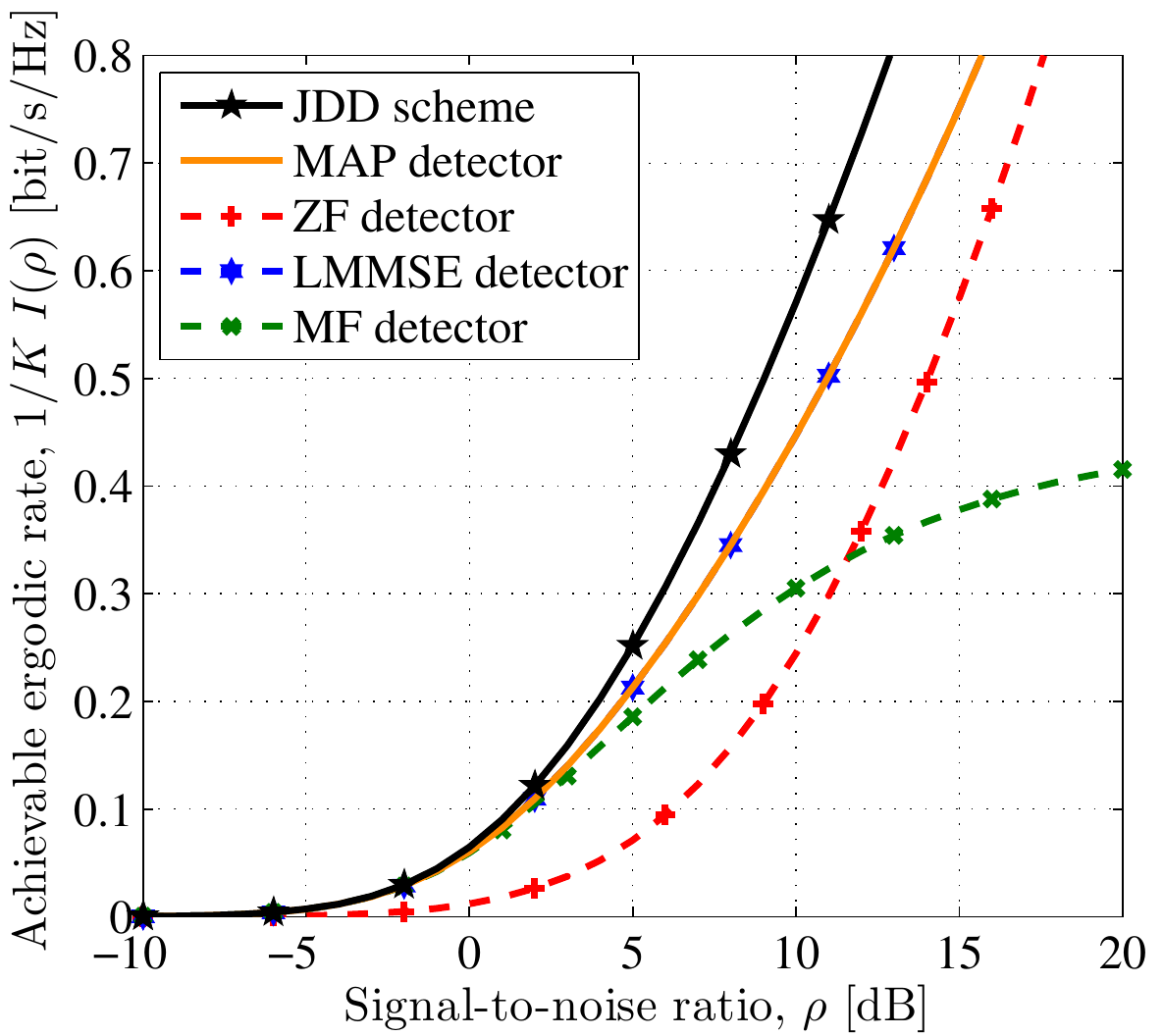}\label{fig:resultMi3HopGaussMfSnr}}\\
		\vspace{-0.25cm}
		\subfigure[QPSK inputs]{\includegraphics[height=7cm,width=8cm]{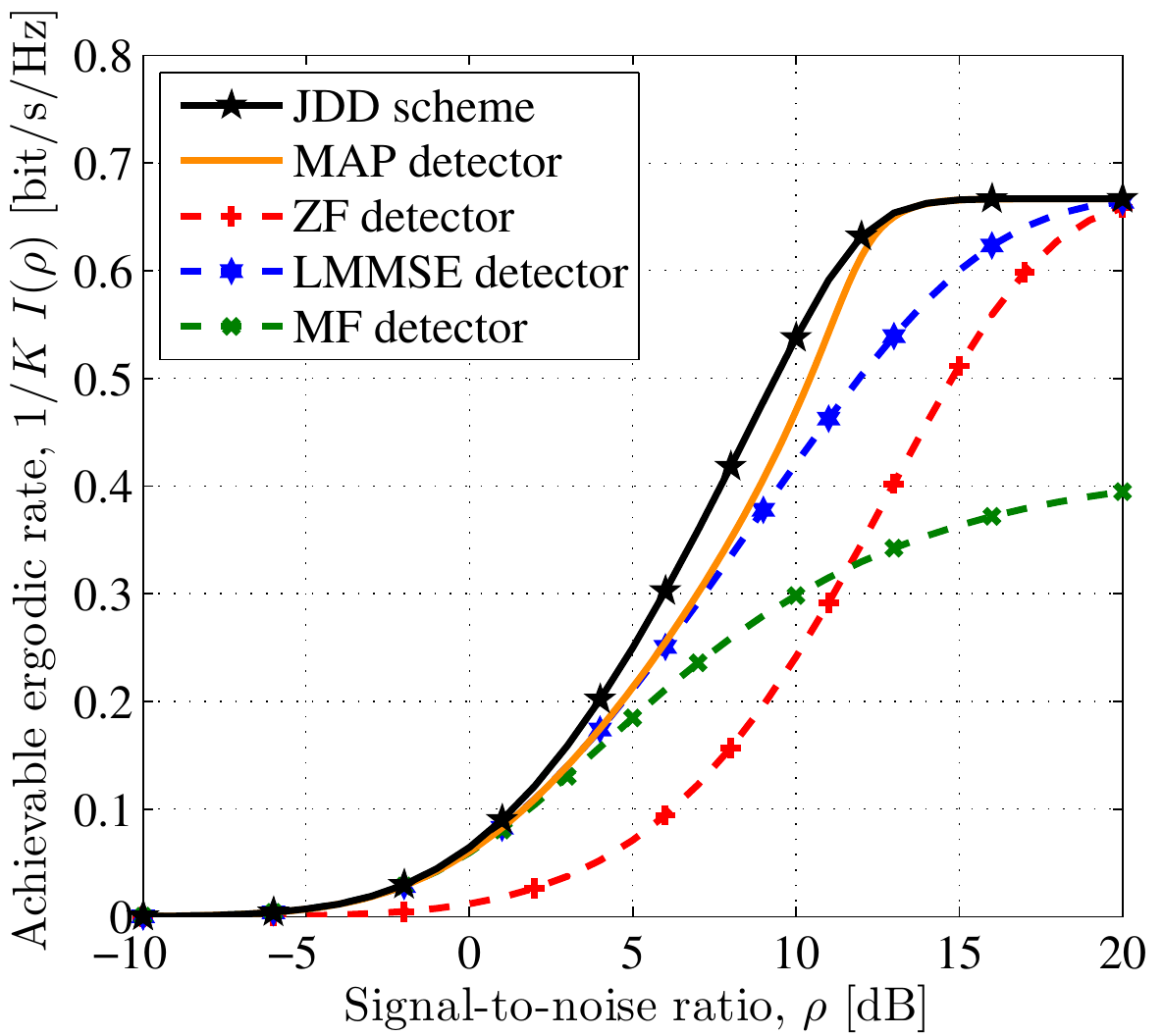}\label{fig:resultMi3HopQpskMfSnr}}
		\caption{Per-dimension achievable rate \vs SNR for an AF relay MIMO channel with $K=3$ hops under Gaussian and QPSK signaling with various detection schemes. A factor of $1/K$ in front of the rate is due to the TDMA protocol. The SNRs for the hops are set to $\rho_1 = \rho$, $\rho_2 = 0.7\rho$ and $\rho_3 = 0.5\rho$. The numbers of antennas at terminals are set to $M_0 = 4$, $M_1 = 6$, $M_2 = 8$ and $M_3 = 12$.}
		\label{fig:resultMi3HopMfSnr}
		\vspace{-0.5cm}
	\end{center}
\end{figure}

To quantify the performance of linear detectors, Fig.~\ref{fig:resultMi3HopMfSnr} plots the achievable ergodic rate as a function of SNR for the network with three hops. To enable illustration of the performance of the ZF detector, we set the numbers of antennas unequal. For instance, we set $M_0=4$, $M_1=6$, $M_2=8$ and $M_3=12$. Meanwhile, the SNRs of the hops are set as $\rho_1 = \rho$, $\rho_2 = 0.7\rho$ and $\rho_3=0.5\rho$. Fig.~\ref{fig:resultMi3HopGaussMfSnr} depicts the performance of various detection schemes under Gaussian channel inputs, while Fig.~\ref{fig:resultMi3HopQpskMfSnr} illustrates the behavior under QPSK inputs. From the figure one can see that, quite expectedly, the JDD scheme outperforms the other depicted detectors. This is because optimal JDD is essentially the best one can do. In addition, from Fig.~\ref{fig:resultMi3HopGaussMfSnr} one notes that for Gaussian signaling performance of the MAP detector matches to that of the LMMSE, which highlights the optimality of the latter in the case of Gaussian signals. 

Another observation is that the MF detector demonstrates near-optimal performance at low SNR and becomes increasingly inefficient as the SNR increases. In turn, the ZF detector demonstrates significant degradation in performance at low SNR, whilst becoming progressively efficient with increasing SNR. The same holds for the case of QPSK inputs, as depicted in Fig.~\ref{fig:resultMi3HopQpskMfSnr}. However, in contrast to Gaussian signaling, where the loss due to separation of decoding (when comparing optimal JDD and MAP detection schemes) grows with SNR, for QPSK constellation this loss vanishes in the high-SNR region due to saturation of the MI in this setup. Moreover, while for the JDD scheme Gaussian signaling always outperforms QPSK, for the MAP detector scheme this is not always the case. Accurate comparison of the two figures reveals that for the small mid-SNR region (roughly between 5 and 10 dB) the QPSK constellation actually performs better than Gaussian signaling under SD. Also, as expected, linear detection schemes (MF, ZF and LMMSE) perform worse.

\subsection{Bit Error Rate}

\begin{figure}
	\centering
	\includegraphics[height=7.2cm,width=8.9cm]{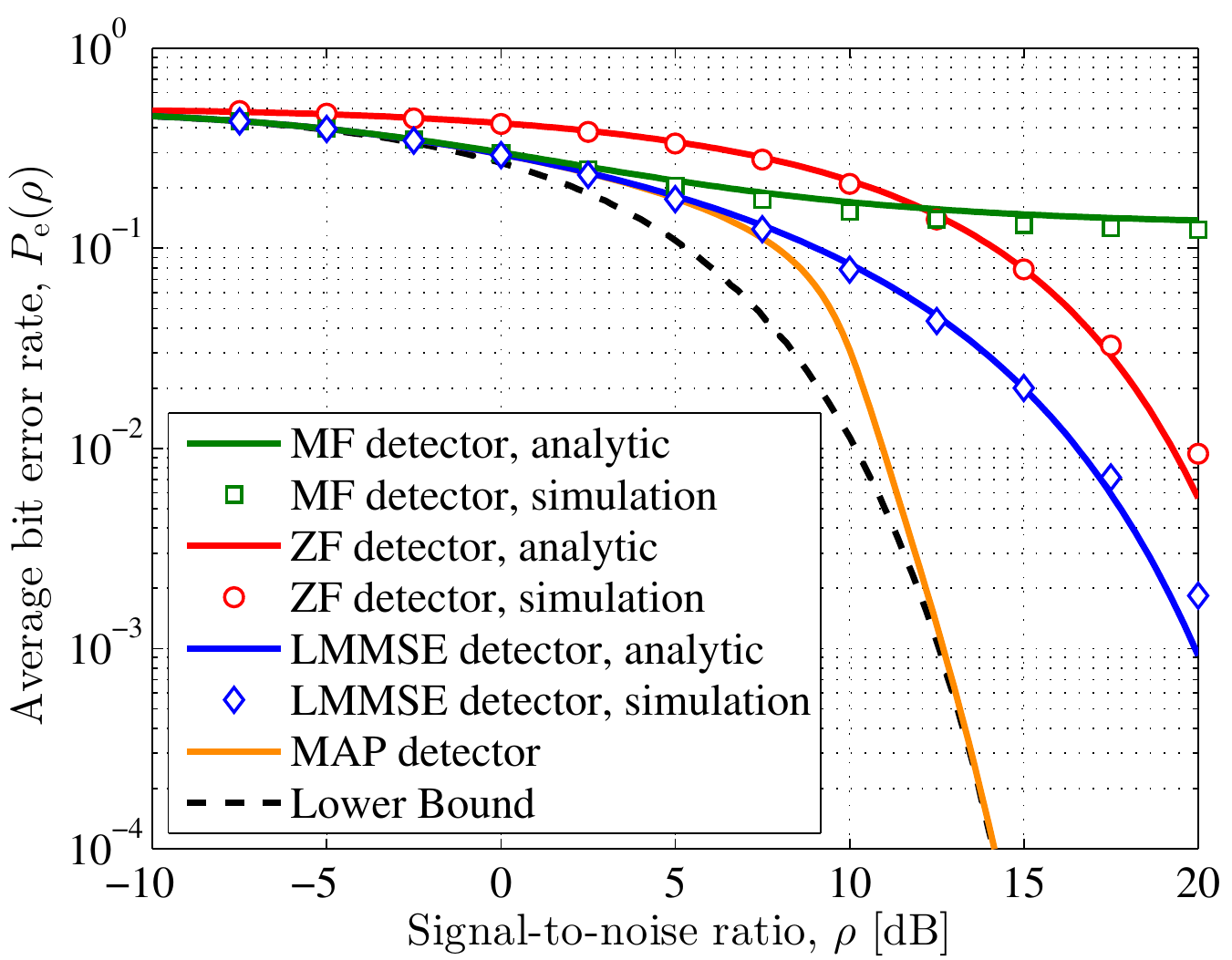}
	\caption{Average uncoded BER \vs SNR for an AF relay MIMO system with $K=2$ hops under QPSK signaling and various detection schemes. Transmit SNRs for two hops are given by $\rho_1 = \rho_2 = \rho$. The numbers of antennas at terminals are set to $M_0 = 24$, $M_1 = 28$ and $M_2 = 36$. Solid curves denote the analytic results, while markers denote the results of Monte Carlo simulations. The black dashed line denotes the lower bound in~\eqref{eqn:lowerBound}.}
	\label{fig:resultBer2HopQpskMcSnr}
\end{figure}

To verify the obtained approximation for BER~\eqref{eqn:berAsymptotic}, we simulate the average  BER of a two-hop AF relay MIMO communication system with QPSK signaling. We fix the numbers of antennas at terminals $M_0 = 24$, $M_1 = 30$ and $M_2 = 36$ and relations between SNRs at each hop $\rho_1 = \rho_2 = \rho$. Fig.~\ref{fig:resultBer2HopQpskMcSnr} plots the results of numerical simulations, averaged over 500 channel realizations. From the figure, we note that the asymptotic result predicts well the behavior of the system at low SNR even for a system with finite (albeit quite large) size. However, as SNR increases the approximation becomes less accurate, and henceforth the diversity order effects, usually visible in that region, are not captured. The reason for this is that at high SNR the analysis becomes much more sensitive to the large-system assumption. It becomes increasingly precise as the system size grows large, while the average BER in such regime tends to a Gaussian Q-function. At the same time, for the sake of comparison Fig.~\ref{fig:resultBer2HopQpskMcSnr} also depicts average BER of the optimal MAP detector. At low SNR the corresponding performance curve follows the upper bound (given by the performance of a linear MMSE detector), whereas at high SNR it tends to the lower bound\footnote{The lower bound is attained with interference-free transmission. That is, since in~\eqref{eqn:channelE2E} the equivalent noise term
	\begin{align*}
		\tilde{\bn} = \sum_{k=1}^{K-1} \bG_k^{K-1} \bn_k + \bn_K
	\end{align*}
	is colored, a whitening filter of form
	\begin{align*}
		\bU = \sqbrc{\sum_{k=1}^{K-1} \bG_k^{K-1}\brc{\bG_k^{K-1}}^{\H} + \bI_{M_K}}^{-1/2},
	\end{align*}
	is applied, which yields a signal with covariance $\E_{\calH}\figbrc{\bC\bC^{\H}}$ and white noise, where $\bC$ is given in~\eqref{eqn:matrixC}.} given by
\begin{align}\label{eqn:lowerBound}
	\pe^{\textrm{LB}}(\brho) = Q\brc{\sqrt{ M_0^{-1}\tr\figbrc{\E_{\calH}\figbrc{\rbC\rbC^{\H}}}}},
\end{align}
where
\begin{align}\label{eqn:matrixC}
	\rbC \triangleq \sqbrc{\bI_{M_K} + \sum_{k=1}^{K-1} \vbG_k^{K-1}\brc{\vbG_k^{K-1}}^{\H}}^{-1/2}\vbG_0^{K-1}.
\end{align}

\begin{figure}
	\centering
	\includegraphics[height=7.2cm,width=8.9cm]{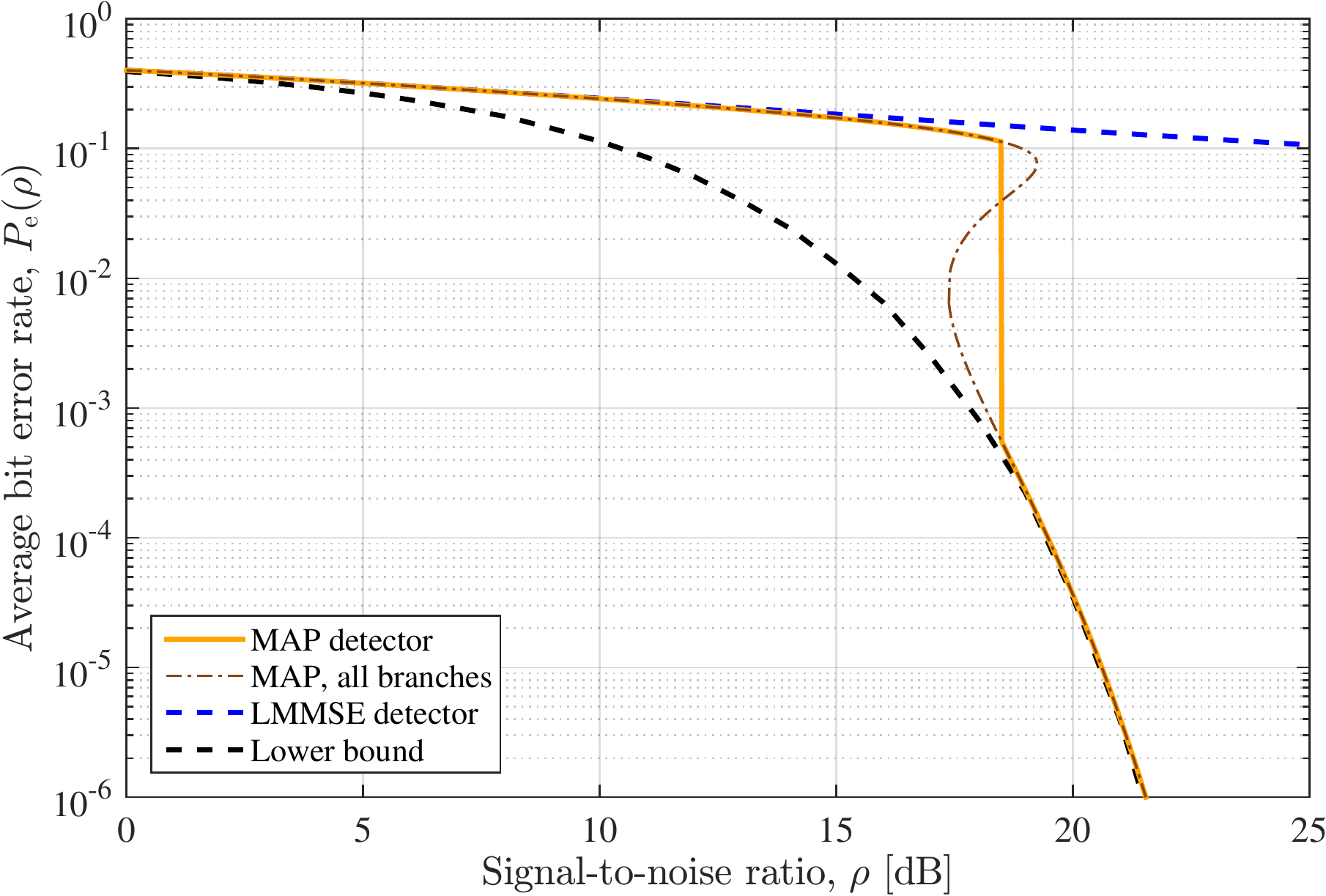}
	\caption{Average uncoded BER \vs SNR for an AF relay MIMO system with $K=3$ hops under QPSK signaling and MAP detection scheme. Transmit SNRs for the hops are given by $\rho_1 = \rho_2 = \rho_3 = \rho$. The numbers of antennas at terminals are set to $M_0 = 10$, $M_1 = 9$, $M_2 = 8$ and $M_3=7$. Solid curve denotes the analytic results, dashed lines denote the performance of LMMSE detector (upper bound) and the lower bound in~\eqref{eqn:lowerBound}. The dash-dotted line denotes the system behavior when taking into account the metastable solution.}
	\label{fig:resultBer3HopQpskOptSnr}
\end{figure}

Meanwhile, Fig.~\ref{fig:resultBer3HopQpskOptSnr} depicts the average BER of a MAP detector for a similar AF relay MIMO scenario but with $K=3$ hops. Here, the numbers of antennas are set to $M_0 = 10$, $M_1 = 9$, $M_2 = 8$ and $M_3=7$, while SNRs of the hops are still equal, \ie, $\rho_1 = \rho_2 = \rho_3 = \rho$. The figure shows a waterfall-like behavior for the performance of the MAP detector.  Namely, the globally stable (true) solution, which minimizes the free energy, instantly switches between the upper and lower bounds at the transition point. At the same time, there exist \emph{metastable} solutions, which are local minimizers for the free energy. Since the detection algorithm does not know the true detection results initially, the system may get trapped in a metastable solution. When taking into account the latter, a Z-shaped curve is obtained for the BER of the MAP detector (depicted by a dash-dotted line in the figure). Note that there is a region where the curve exhibits an increase in BER with increasing the SNR. The true system BER, however, never gets to that point and instead manifests a sharp transition when reaching the transition point. This behavior is somewhat similar to that of Fig.~\ref{fig:resultMi3Hop8PskSdSnr}, also indicating the occurrence of a phase transition.

\section{Conclusions}\label{sec:conclusion}

In this paper, we have developed a framework for the asymptotic performance analysis of a $K$-hop AF relay MIMO system with arbitrary $K$ and discrete channel inputs. The framework captures the effects of separation between detection and decoding, as well as suboptimality of linear detectors. More precisely, we have evaluated the performance of the system under separate detection in terms of achievable ergodic data rate, as well as average bit error rate, in the limit where antenna arrays grow large without bounds. The main result states that the $K$-hop AF relay MIMO channel with the GPME detector at the destination terminal decouples into a bank of per-stream scalar channel with a GPME detector front end. Comparing to Monte Carlo simulations, it has been confirmed that the results provide an accurate approximation for a finite-sized system at low and moderate SNR. It has been further shown \via a numerical example that the number of hops in a multi-hop AF network might have a significant impact on the system performance and might be properly adjusted using the present results. Moreover, the individually optimal detection scheme is shown to experience a phase transition at certain SNR values. The obtained compact expressions may also be useful for the design of coding schemes improving the system performance. The presented results are potentially extendable to more sophisticated channel models of interest (\eg, Kronecker model or Rician fading) allowing for further performance optimization as in \cite{Girnyk-etal-twc2014}.  Another practical problem of interest would be to consider the case of covariance mismatched decoding \cite{Vehkapera-etal-trcom2015}, where the receiver does not know the instantaneous realizations of the intermediate channels.

\appendices

\section{Derivation of Claim~\ref{thm:freeEnergyAsymptotic}}
\label{sec:proofFreeEnergy}

Recall the expression of the partition function given in~\eqref{eqn:partitionFunction}
\begin{equation}
\label{eqn:partitionFunctionRecall}
Z(\by, \calH) \triangleq \E_{\bx,\calN}\figbrc{\frac{1}{\pi^{M_K}} \e^{-\norm{\by - \bG_0^{K-1} \bx - \sum_{k=1}^{K-1} \bG_k^{K-1} \bn_k}^2}}.
\end{equation}
The free energy in~\eqref{eqn:freeEnergyIdentity} can thus be rewritten as
\begin{subequations}
	\begin{align}
		\calF =& -\frac{1}{M_0}\E_{\by,\calH} \ln Z(\by, \calH)\label{eqn:relationEntropyPatritionFunction}\\
		=& - \frac{1}{M_0}\lim_{u \rightarrow 0^+} \frac{\partial}{\partial u} \ln \E_{\by,\calH} \figbrc{Z^u(\by,\calH)}. \label{eqn:identity}
	\end{align}
\end{subequations}
Since for real-valued $u$ computing $\E_{\by,\calH}\{Z^u(\by,\calH)\}$ is very difficult, we make a non-rigorous assumption of \emph{replica continuity}. That is, it is postulated that the $u$th moment of the partition function can be first evaluated for integer $u$ and then generalized to real-valued $u$ assuming analytic continuation\footnote{This step constitutes one of the major problems with the replica method having been unproved rigorous yet. The validity of this assumption is an ongoing problem in mathematical physics. For detailed discussion see~\cite{tanaka2007moment}.}. We proceed with the so-called \emph{replication}, \ie, we introduce $u$ replicas of the postulated channel as follows
\begin{align}
	&\!\!\E_{\by,\calH}\! \figbrc{\!Z^u(\by,\calH)\!} \!=\! \E\! \figbrc{ \!\int\!\! \frac{1}{\pi^{M_K}}\e^{\!-\norm{\by - \bG_0^{K\!-\!1} \bxa{0} - \sum\limits_{k=1}^{K-1} \bG_k^{K\!-\!1} \bna{0}_k}^2}\!\! \right.\nonumber\\
		&\times \left.\frac{1}{(\pi\sigma^2)^{M_K}}\!\prod_{a=1}^u\! \e^{\!-\norm{\by - \bG_0^{K\!-\!1} \bxaPr{a} - \sum\limits_{k=1}^{K-1} \bG_k^{K\!-\!1} \bnaPr{a}_k}^2}\!\! \d\by\!}, \label{eqn:replication}
\end{align}
where $\bxaPr{a}$ and $\bnaPr{a}_k$ denote the $a$th replica vectors that are assumed to be i.i.d., while $\bxa{0}$ and $\bna{0}_k$ represent the original signal vector and the noise vector at the $k$th hop. For ease of exposition, we group the corresponding vectors into $\bX = [\bxaT{0}, \bxaPrT{1},\ldots, \bxaPrT{u}]^{\T} \in \complex{M_0(u+1)}$ and $\bN_k = [\bnaT{0}_k, \bnaPrT{1}_k,\ldots, \bnaPrT{u}_k]^{\T} \in \complex{M_k(u+1)}$.

In~\eqref{eqn:replication}, the averaging should be performed over all possible channel inputs $\bX$, channel gains $\calH$ and noise realizations $\calN$. According to the Fubini theorem~\cite[Theorem 18.3]{billingsley1995probability}, provided that the expectation in~\eqref{eqn:replication} exists, the multiple integral can be computed \via repeated integrals. In other words, averaging over the channel matrices and noise vectors of all hops can be done iteratively hop-by-hop. That is, at each hop, we average out the randomness of the corresponding channel matrix and the noise vector, while keeping the variables related to other hops fixed. To do this, define the following set of vectors
\begin{subequations}
	\begin{align}
		\bva{0}_1 &\triangleq \bH \bxa{0} + \bna{0}_1 \in \mathds{C}^{M_1},\\
		\bva{a}_1 &\triangleq \bH \bxaPr{a} + \bnaPr{a}_1 \in \mathds{C}^{M_1},
	\end{align}
\end{subequations}
for all $a \in\{1,\ldots,u\}$ containing the randomness of the first hop in the network. For each subsequent hop $k$ vector $\bva{a}_k$ is recursively defined in terms of vector $\bva{a}_{k-1}$, containing the randomness of first $k-1$ hops, as follows
\begin{equation}
\bva{a}_k = \bH_k \bva{a}_{k-1} \in \mathds{C}^{M_{k}}.
\end{equation}
For each $k$, stack these vectors as $\bV_k \triangleq [\bvaT{0}_k, \ldots, \bvaT{u}_k ]^{\T} \in \mathds{C}^{M_k(u+1)}$. For $k\in\{2,\ldots,K-1\}$, conditioned on $\{\bH_1, \ldots, \bH_{k-1}\}$ and $\{\bn^{(a)}_1, \ldots, \bn^{(a)}_{k-1}\}, \; a \in \{1, \ldots, u\}$, vector $\bV_k$ is a complex Gaussian random vector (\vide~\cite{tanaka2002statistical}) with the covariance matrix given by
\begin{equation}
\label{eqn:matrixCov}
\bK_k = (\bQ_k + \bE) \otimes\bI_{M_k} \in \complex{M_k(u+1)\times M_k(u+1)},
\end{equation}
where
\begin{equation}
\label{eqn:matrixQ}
[\bQ_k]_{a,b} \triangleq \frac{\rho_k \beta_{k-1}}{M_{k-1}} \bvaH{b}_{k-1} \bva{a}_{k-1} \in \complex{(u+1)\times(u+1)}.
\end{equation}
and $\bE \triangleq
\sqbrc{\begin{smallmatrix}
	1 & \bzero{1\times u}\\
	\bzero{u\times 1} & \sigma^2 \bI_u
	\end{smallmatrix}}\in\real{(u+1)\times(u+1)}$.

\begin{figure*}[t!]
	\normalsize
	\setcounter{tempeqncnt}{\value{equation}}
	\setcounter{equation}{50}
	\begin{align}
		\label{eqn:T1}
		\TU_{1}(\calQ_{1},\til{\calQ}_{1}) =&\; \alp_{0,K}\ln\brc{1+\frac{u}{\sigma^2}} + \alp_{0,K} \ln\det\brc{\bI_{u+1} + \bQ_{K}\bSigma} - \frac{1}{M_0}\ln \E_{\bX} \figbrc{\exp\sqbrc{\rho_{1}\beta_{0} \bX^{\H}(\til{\bQ}_{1}\otimes\bI_{M_{0}})\bX}} \nonumber\\
		& +\! \sum_{k=1}^{K-1} \alp_{0,k}\ln\det\sqbrc{\bI_{u+1} - \rho_{k+1}\beta_{k}\til{\bQ}_{k+1}\brc{\bQ_{k}+\bE}} + \sum_{k=1}^{K} \alp_{0,k-1} \tr\{\bQ_k \til{\bQ}_{k}\} + u\alp_{0,K} \ln\pi\sigma^2
	\end{align}
	\setcounter{equation}{52}
	\hrulefill
	\begin{align}\label{eqn:tFunctionSimplified}
		\TU_1&(\calQ_1,\tilcalQ_1) = (u-1) \alp_{0,K} \ln\brc{1+\frac{\pk{K}-\qk{K}}{\sigma^2}} + \alp_{0,K} \ln \sqbrc{1 + \frac{\pk{K}-\qk{K}}{\sigma^2} + \frac{u}{\sigma^2}(1+\rk{K}-\mk{K}-\mk{K}^*+\qk{K})}\nonumber\\
		&+ \alp_{0,K} u\ln\pi\sigma^2 +\sum_{k=1}^{K} \alp_{0,k-1}\brc{\tilrk{k} \rk{k} + u \tilmk{k} \mk{k} + u \tilmk{k}^* \mk{k}^* + u \tilpk{k} \pk{k} + u (u-1) \tilqk{k} \qk{k}} \nonumber\\
		&+(u-1)\sum_{k=1}^{K-1} \alp_{0,k}\ln\sqbrc{1-\beta_{k}\rho_{k+1} (\tilpk{k+1}-\tilqk{k+1})\brc{\sigma^2 + \pk{k}-\qk{k}}}\nonumber\\
		& + \sum_{k=1}^{K-1}\alp_{0,k}\ln \Big(1-\beta_k\rho_{k+1}\sqbrc{u\tilmk{k+1}^* \mk{k} + u\tilmk{k+1} \mk{k}^* + (\tilpk{k+1} +(u-1)\tilqk{k+1})\brc{\sigma^2 + \pk{k} +(u-1)\qk{k}} + \tilrk{k+1}(1+\rk{k})}\nonumber\\
		&\qquad\qquad\qquad+\beta_k^2\rho_{k+1}^2\sqbrc{u|\tilmk{k+1}|^2 - (1+\rk{k})\brc{\sigma^2 + \pk{k} +(u-1)\qk{k}}}\sqbrc{u|\mk{k}|^2 - \tilrk{k+1}(\tilpk{k+1} +(u-1)\tilqk{k+1})}\Big)\nonumber\\
		&- \ln\frac{\eta_1}{\pi}-\frac{1}{M_0}\ln \int \E_{\vbx} \figbrc{\e^{- \eta_1 \norm{\rbz - \sqrt{\rho_1\beta_0} \rbx}^2}\e^{\rho_1\beta_0 \phi_1\rbx^{\H}\rbx}} \sqbrc{\E_{\vbx'} \figbrc{\e^{-\xi_1 \norm{\rbz - \sqrt{\rho_1\beta_0}\rbx'}^2} \e^{\xi_1 \rbz^{\H}\rbz + \rho_1\beta_0 \psi_1 \rbx'^{\H}\rbx'}}}^u \d \rbz
	\end{align}
	\setcounter{equation}{\value{tempeqncnt}}
	\addtocounter{topeqncnt}{1}
	\hrulefill
	\vspace{-0.5cm}
\end{figure*}

Starting from hop $K$, we proceed in a similar way as in~\cite{guo2005randomly},~\cite{tanaka2002statistical} and write~\eqref{eqn:replication} as
\begin{equation}
\frac{1}{M_0}\!\ln\E_{\by,\calH} \figbrc{Z^u(\by,\calH) }\! =\! \frac{1}{M_0}\ln\!\int\! \e^{M_K\Gu_K(\bQ_K)} \d\muu_K(\bQ_K),
\end{equation}
where we have omitted all the vanishing terms,
\begin{align}\label{eqn:gFunction}
	\Gu_K(\bQ_K) \triangleq&  -\ln (u+1) - \ln \det \brc{\bI_{M_K(u+1)} + \bQ_K \bSigma},
\end{align}
and the probability measure of $\bQ_K$ is given by
\begin{equation}
\muu_K(\bQ_K)\! =\! \E\figbrc{\!\prod_{a,b=0}^u\!\! \delta\big(\rho_{K}\beta_{K\!-\!1} \bvaH{b}_{K\!-\!1}\bva{a}_{K-1}\! - M_{K\!-\!1}[\bQ_K]_{a,b}\big)\!\!}.
\end{equation}

The moment-generating function (MGF) induced by $\muu_K(\bQ_K)$ is given by
\begin{equation}
\label{eqn:mgf}
\MU_K(\til{\bQ}_K) = \E_{\bV_{K-1}}\figbrc{\e^{\rho_K \beta_{K-1}\bV_{K-1}^{\H}(\til{\bQ}_K \otimes \bI_{M_{K-1}}) \bV_{K-1}}},
\end{equation}
which yields the rate function
\begin{equation}
\label{eqn:rateFunctionK}
\IU_K(\bQ_K) = \underset{\til{\bQ}_K}{\max}\figbrc{\tr \{\til{\bQ}_K \bQ_K\} - \frac{1}{M_{K-1}}\ln \MU_{K}(\til{\bQ}_{K})}.
\end{equation}
Hence, by the G\"{a}rtner-Ellis theorem~\cite[Theorem 2.3.6]{dembo1998large}, in the LSL
\begin{align}
	&\frac{1}{M_0}\ln \E_{\by,\calH} \figbrc{Z^u(\by,\calH)} \nonumber\\
	&- \underset{\bQ_K}{\max}\figbrc{\alp_{0,K}\Gu_K(\bQ_K) - \alp_{0,K-1}\IU_K(\bQ_K) } \to 0.
\end{align}
Combining all together, we get at hop $K$
\begin{equation}
\frac{1}{M_0}\ln\E_{\by,\calH} \figbrc{Z^u(\by,\calH)} =  \underset{\calQ_{K}}{\min}\; \underset{\til{\calQ}_{K}}{\max} \figbrc{\TU_{K}(\calQ_{K}^K,\til{\calQ}_{K}^K)},
\end{equation}
where $\calQ_{i}^j \triangleq \{\bQ_{i},\ldots,\bQ_{j}\}$, $\til{\calQ}_{i}^j \triangleq \{\til{\bQ}_K,\ldots,\til{\bQ}_j\}$, and
\begin{align}
	\label{eqn:expressionTK}
	\TU_{K}&(\calQ_{K}^K,\til{\calQ}_{K}^K) = \alp_{0,K-1} \tr\{\bQ_K \til{\bQ}_{K}\} - \frac{1}{M_0}\ln \MU_{K}(\til{\bQ}_{K})\nonumber\\
	& + \alp_{0,K}\ln\pi(u+1) + \alp_{0,K}\ln\det(\bI_{u+1} + \bQ_{K}\bSigma),
\end{align}
where $\bSigma \triangleq \bI_{u+1} - \frac{1}{u+1}\bone{u+1}\boneT{u+1}$.

We now need to evaluate the second term in~\eqref{eqn:expressionTK} following the same procedure as above. Namely, we rewrite it, omitting the vanishing terms, as
\begin{equation}
\label{eqn:mgfRewrite}
\frac{1}{M_0}\ln\MU_K\!(\til{\bQ}_K)\! =\! \frac{1}{M_0}\ln\!\int\!\! \e^{M_K\Gu_{K-1}\!(\bQ_{K-1})} \d\muu_{K-1}(\bQ_{K-1}),
\end{equation}
where, using the Gaussian integral, we can obtain
\begin{align}
	\label{eqn:mgfRewrite}
	\Gu_{K-1}&(\bQ_{K-1})\nonumber\\
	&= \alp_{0,K-1}\ln\det\sqbrc{\bI_{u+1} - \rho_{K}\beta_{K-1}\til{\bQ}_{K}\brc{\bQ_{K-1}+\bE}}.
\end{align}
Now, proceeding with the same steps as before, we arrive at
\begin{equation}
\frac{1}{M_0}\ln\MU_K(\til{\bQ}_K) -  \underset{\calQ_{K-1}}{\min}\; \underset{\til{\calQ}_{K-1}}{\max} 
\Big\{\TU_{K-1}(\calQ_{K-1}^K,\til{\calQ}_{K-1}^K) \Big\} \to 0,
\end{equation}
where
\begin{align}
	\label{eqn:expressionTK1}
	\TU_{K-1}&(\calQ_{K-1}^K,\til{\calQ}_{K-1}^K) =  - \frac{1}{M_0}\ln \MU_{K-1}(\til{\bQ}_{K-1})\nonumber\\
	+&\; \alp_{0,K-1} \tr\{\bQ_K \til{\bQ}_{K}\} + \alp_{0,K-2} \tr\{\bQ_{K-1} \til{\bQ}_{K-1}\}\nonumber\\
	+&\; \alp_{0,K-1}\ln\det\sqbrc{\bI_{u+1} - \rho_{K}\beta_{K-1}\til{\bQ}_{K}\brc{\bQ_{K-1}+\bE}}\nonumber\\
	+&\; \alp_{0,K}\ln\pi(u+1) + \alp_{0,K}\ln\det(\bI_{u+1} + \bQ_{K}\bSigma).
\end{align}
Proceeding with the same procedure for all the hops, we can show that for $k \in \{2,\ldots,K\}$ the corresponding log-MGF term can be written as
\begin{align}
	\label{eqn:induction}
	-&\frac{1}{M_0}\ln \MU_{k}(\til{\bQ}_{k})\nonumber\\
	= &\;\alp_{0,k-1}\ln\det\sqbrc{\bI_{u+1} - \rho_{k-1}\beta_{k-2}\til{\bQ}_{k}\brc{\bQ_{k-1}+\bE}}\nonumber\\
	&- \frac{1}{M_0}\ln \MU_{k-1}(\til{\bQ}_{k-1})  + \alp_{0,k-2} \tr\{\bQ_{k-1} \til{\bQ}_{k-1}\}.
\end{align}
Thus, we iteratively evaluate $\MU_{k}(\til{\bQ}_{k}), \; \forall k$ and arrive at
\begin{equation}
\label{eqn:freeEnergyFinal}
\calF = - \frac{1}{M_0}\lim_{u \rightarrow 0^+} \frac{\partial}{\partial u} \;\underset{\calQ_{1}}{\min}\; \underset{\til{\calQ}_{1}}{\max} \figbrc{\TU_{1}(\calQ_{1},\til{\calQ}_{1})},
\end{equation}
where $\TU_{1}(\calQ_{1},\til{\calQ}_{1})$ reads as~\eqref{eqn:T1} at the top of the next page.
\addtocounter{equation}{1}

Finding the fixed point of~\eqref{eqn:freeEnergyFinal} is a complicated task and may not be realizable directly. Hence, a simplifying \emph{replica-symmetry} (RS) assumption is made in order to proceed. Namely, all $k$ matrices $\bQ_k$ and $\til{\bQ}_k$ are postulated to have the following structure
\begin{subequations}
	\label{eqn:rsAnsatz}
	\begin{align}
		\bQ_k =&\;
		\begin{pmat}[{|.}]
			\rk{k} & \mk{k} \bone{u}^{\T}\cr\-
			\mk{k}^* \bone{u} & (\pk{k}-\qk{k})\bI_u + \qk{k} \bone{u}\bone{u}^{\T}\cr
		\end{pmat}\!,\\
		\tilrbQ =&\;
		\begin{pmat}[{|.}]
			\tilrk{k} & \tilmk{k} \bone{u}^{\T}\cr\-
			\tilmk{k}^* \bone{u} & (\tilpk{k}-\tilqk{k})\bI_u + \tilqk{k} \bone{u}\bone{u}^{\T}\cr
		\end{pmat}\!.
	\end{align}
\end{subequations}

The assumption has been widely accepted in the literature following the reasoning that the physics of the whole system should not depend on the artificially introduced replica indices.\footnote{It is noteworthy that there have been reported cases where \emph{replica-symmetry breaking} occurs~\cite{muller2008vector},~\cite{zaidel2012vector} and the RS-based approach fails. In such cases, one has to carry out the calculations using an RSB ansatz, which leads to much more involved mathematics.}

With the RS assumption,~\eqref{eqn:T1} is simplified \via the Gaussian linearization (based on the Hubbard-Stratonovich transform~\cite{Hubbard1959},~\cite{stratonovich1957method}), as carried out in~\cite{guo2005randomly}, to~\eqref{eqn:tFunctionSimplified} on the top of the page, where $\eta_1 \triangleq \frac{|\tilmk{1}|^2}{\tilqk{1}}$, $\phi_1 \triangleq \tilrk{1}$, $\xi_1 \triangleq \tilmk{1}^*$ and $\psi_1 \triangleq \tilmk{1}^* + \tilpk{1}-\tilqk{1}$.
\addtocounter{equation}{1}

The last part of~\eqref{eqn:tFunctionSimplified} is thus going to be associated with the two fixed Gaussian scalar channels below
\begin{subequations}
	\begin{align}
		\vz =&\; \sqrt{\beta_0\rho}\; \vx + \frac{\vw}{\sqrt{\eta_1}},\\
		\vz =&\; \sqrt{\beta_0\rho}\; \vx' + \frac{\vw'}{\sqrt{\xi_1}},
	\end{align}
\end{subequations}
where $\vw,\vw' \sim \calC\calN (0,1)$.

Now, to find the saddle point in~\eqref{eqn:freeEnergyFinal}, we have to take the derivatives of $\TU_1(\calQ,\tilcalQ)$ w.r.t. to all the $8K$ parameters. We find that $\tilrk{k} = 0$ and $\tilmk{k}^* = \tilmk{k}, \; \forall k$, and $\tilpk{1} - \tilqk{1} = -\tilmk{1}$. Furthermore, we get
\begin{subequations}
	\begin{align}
		\rk{1}-\mk{1}-\mk{1}^*+\qk{1} =&\; \beta_0\rho \E_{\vz,\vx} \figbrc{|x - \mmse{x'}|^2},\\
		\pk{1}-\qk{1} =&\; \beta_0\rho \E_{\vz,\vx'} \figbrc{|x' - \mmse{x'}|^2},
	\end{align}
\end{subequations}
thus obtaining the set of fixed-point equations~\eqref{eqn:fpEqs}. To further evaluate the free energy, we take the derivative of $\TU_1(\calQ_1,\tilcalQ_1)$ w.r.t. $u$ at $u \to 0^+$ and--keeping in mind that $\rk{1} = \beta_0\rho_1$ and $\rk{k} = \beta_{k-1}\rho_k(1+\beta_{k-2}\rho_{k-1})$--obtain~\eqref{eqn:freeEnergy}, where we have denoted $\eta_k \triangleq \frac{\tilmk{k}^2}{\tilqk{k}}$, $\xi_k \triangleq \tilmk{k}$, $\epsk{k}\triangleq\rk{k}-\mk{k}-\mk{k}^*+\qk{k}$ and $\nuk{k}\triangleq\pk{k}-\qk{k}$, for all $k\in\{1,\ldots,K\}$.

\section{Derivation of Claim~\ref{thm:decoupling}}
\label{sec:proofDecoupling}

Consider the channel input, the postulated input and the output of the GPME for $m, \; m = 1,\ldots,M$. We want to evaluate the joint moments of the joint distribution of $(x_{m}, x',\mmse{x'_m}_q)$. For non-negative integers $\ir$, $\ii$, $\jr$, $\ji$, with $\br,\bi\in\{1,\ldots,u\}, \; \br\neq\bi$, let $\calAr$ and $\calAi$ be disjoined subsets of $\{1,\ldots,u\}\setminus\{\br,\bi\}$ with cardinalities $\lr$ and $\li$, respectively. Define a function
\begin{align}
	g(\rbX&) \triangleq \sum_{m=1}^M (\Re\{x_m^{(0)}\})^{\ir} (\Im\{x_m^{(0)}\})^{\ii} (\Re\{x_m^{(\br)}\})^{\jr} \nonumber\\
	&\times  (\Im\{x_m^{(\bi)}\})^{\ji} \prod_{\ar\in \calAr} \Re\{x_m^{(\ar)}\} \prod_{\ai\in\calAi} \Im\{x_m^{(\ai)}\}.
\end{align}
Let us furthermore introduce an infinitesimal perturbation into the Hamiltonian in~\eqref{eqn:partitionFunctionRecall}, so that
\begin{align}\label{eqn:partitionFunctionNew}
	&\Ztilu(\rby,\!\calH,\!\rbx;\omega\!)\!  \nonumber\\
	&\triangleq \!\E_{\bX,\calH} \Bigg\{\! \frac{\e^{\omega g(\rbX)}}{\pi^{M_K}(\pi\sigma^2)^{u M_K}} \!\!\int\! \e^{\!-\norm{\by - \bG_0^{K\!-\!1} \bxa{0} - \sum_{k=1}^{K-1} \bG_k^{K\!-\!1} \bna{0}_k}^2}\!\! \nonumber\\
	&\qquad\quad\times  \prod_{a=1}^u \e^{-\norm{\by - \bG_0^{K\!-\!1} \bxaPr{a} - \sum_{k=1}^{K-1} \bG_k^{K\!-\!1} \bnaPr{a}_k}^2} \d \by \Bigg\},
\end{align}
is the partition function of a related large system. Here we emphasize that for $\omega = 0$, we have exactly $\Ztilu(\rby,\calH,\rbx;\omega) = \Ztilu(\rby,\calH)$ from~\eqref{eqn:partitionFunctionRecall}. Define then the generalized free energy\footnote{Strictly speaking, this object is not free energy due to its sign. Nevertheless, defined as it is, the quantity provides us with the joint moments of the joint distribution of $(\vx_{m},\vx'_m,\mmse{\vx'_m}_q)$.} as follows
\begin{align}\label{eqn:freeEnergyNew}
	\til \calF =&\; \frac{1}{M_0} \lim_{u \to 0^+} \frac{\partial}{\partial \omega} \ln \E_{\vby,\calH,\vbx} \Big\{\Ztilu(\rby,\calH,\rbx;\omega)\Big\}\Big|_{\omega = 0},
\end{align}
providing exactly the joint moments of interest.

We proceed with exactly the same steps as in the previous proof, \ie,
\begin{align}
	\frac{1}{M_0}\ln\E &\figbrc{\Ztilu(\rby,\calH,\rbx;\omega)}\nonumber\\
	&\qquad= \frac{1}{M_0}\ln \int \e^{M_K\Gu_K(\bQ_K)} \d \muu_K(\bQ_K; \omega)
\end{align}
without the vanishing constants, where
\begin{align}
	&\muu_K(\bQ_K; \omega)\nonumber\\
	& = \E\figbrc{\e^{\omega g(\rbX)}\!\!\prod_{a,b=0}^u\!\! \delta\big(\rho_{K}\beta_{K\!-\!1} \bvaH{b}_{K\!-\!1}\bva{a}_{K-1}\! - M_{K\!-\!1}[\bQ_K]_{a,b}\big)\!\!}
\end{align}
and function $\Gu_K(\bQ_K)$ is obtained as
\begin{align}
	\label{eqn:gFunction}
	\Gu_K(\bQ_K)\! =\! - u\ln\pi\sigma^2 \!\!-\! \ln\Big(1\!+\!\frac{u}{\sigma^2}\Big)\! +\!\ln\det \big(\bI_{u+1}\!+\! \bQ_K\bSigma\big)\!.
\end{align}
The corresponding MGF for $\muu_K(\bQ_K; \omega)$ is given by
\begin{align}
	\label{eqn:mgf}
	&\MU_K(\til{\bQ}_K;\omega)\nonumber\\
	& = \E_{\bV_{K-1}}\figbrc{\e^{\omega g(\rbX)}\e^{\rho_K \beta_{K-1}\bV_{K-1}^{\H}(\til{\bQ}_K \otimes \bI_{M_{K-1}}) \bV_{K-1}}},
\end{align}
which yields the rate function
\begin{equation}
\label{eqn:rateFunctionK}
\IU_K(\bQ_K;\omega) = \underset{\til{\bQ}_K}{\max}\Big\{\tr \{\til{\bQ}_K \bQ_K\} - \frac{1}{M_{K-1}}\ln \MU_{K}(\til{\bQ}_{K};\omega)\!\Big\}\!.
\end{equation}
Again, by the G\"{a}rtner-Ellis theorem, in the LSL
\begin{align}
	&\frac{1}{M_0}\ln \E_{\by,\calH} \figbrc{Z^u(\by,\calH;\omega)}\nonumber\\
	& - \underset{\bQ_K}{\max}\figbrc{\alp_{0,K}\Gu_K(\bQ_K) - \alp_{0,K-1}\IU_K(\bQ_K; \omega)}\to 0.
\end{align}
Combining all together, as before, we get at hop $K$
\begin{align}
	\frac{1}{M_0}\! \ln\E_{\by,\calH} \figbrc{\!\Ztilu(\by,\calH;\omega)\!}\! =\!  \underset{\calQ_{K}}{\min}\; \underset{\til{\calQ}_{K}}{\max} \figbrc{\TU_{K}\!(\calQ_{K}^K,\til{\calQ}_{K}^K;\omega)\!}\!,
\end{align}
where
\begin{align}
	&\TU_{K}(\calQ_{K}^K,\til{\calQ}_{K}^K;\omega) = \alp_{0,K-1} \tr\{\bQ_K \til{\bQ}_{K}\} +\alp_{0,K} \ln\pi(u\!+\!1)\nonumber\\
	&\! - \frac{1}{M_0}\ln \MU_{K}(\til{\bQ}_{K};\omega)  + \alp_{0,K} \ln\det(\bI_{u\!+\!1} \!+\! \bQ_{K}\bSigma).
\end{align}

Here we notice that the term $\e^{\omega g(\rbX)}$ is present only in the log-MGF term $\ln \MU_{k}(\til{\bQ}_{k};\omega)$ above for all $k$. Thus, as before, we conclude that
\begin{align}
	-&\frac{1}{M_0}\ln \MU_{k}(\til{\bQ}_{k};\omega) \nonumber\\
	=&\;\alp_{0,k-1}\ln\det\sqbrc{\bI_{u+1} - \rho_{k-1}\beta_{k-2}\til{\bQ}_{k}\brc{\bQ_{k-1}+\bE}} \nonumber\\
	& - \frac{1}{M_0}\ln \MU_{k-1}(\til{\bQ}_{k-1};\omega) + \alp_{0,k-2} \tr\{\bQ_{k-1} \til{\bQ}_{k-1}\},
\end{align}
and hence we have
\begin{equation}
\til \calF = - \frac{1}{M_0}\lim_{u \rightarrow 0^+} \frac{\partial}{\partial u} \underset{\calQ_{1}}{\min}\; \underset{\til{\calQ}_{1}}{\max} \figbrc{\TU_{1}\!(\calQ_{1},\til{\calQ}_{1};\omega)\!}\!\bigg|_{\omega = 0},
\end{equation}
where $\TU_{1}(\calQ_{1},\til{\calQ}_{1};\omega)$ is given by
\begin{align}
	&\TU_{1}(\calQ_{1},\til{\calQ}_{1};\omega) = u\alp_{0,K} \ln\pi\sigma^2 + \alp_{0,K}\ln\brc{1+\frac{u}{\sigma^2}}\nonumber\\
	& + \alp_{0,K} \ln\det\brc{\bI_{u+1} + \bQ_{K}\bSigma}+ \sum_{k=1}^{K} \alp_{0,k-1} \tr\{\bQ_k \til{\bQ}_{k}\} \nonumber\\
	& +\! \sum_{k=1}^{K-1} \alp_{0,k}\ln\det\sqbrc{\bI_{u+1} - \rho_{k+1}\beta_{k}\til{\bQ}_{k+1}\brc{\bQ_{k}+\bE}}\nonumber\\
	& - \frac{1}{M_0}\ln \E_{\bX} \figbrc{\e^{\omega g(\rbX)}\e^{\rho_{1}\beta_{0} \bX^{\H}(\til{\bQ}_{1}\otimes\bI_{M_{0}})\bX}}.
\end{align}
After adopting the RS assumption, we get $\TU_{1}(\calQ_{1},\til{\calQ}_{1};\omega)$ similar to~\eqref{eqn:tFunctionSimplified} with the exception of the last term, which now reads
\begin{align}
	&\frac{1}{M_0}\ln \int \E_{\bX} \Bigg\{\e^{\omega g(\rbX)}\e^{- \eta_1 \norm{\rbz - \sqrt{\rho_1\beta_0} \rbx}^2}\e^{\rho_1\beta_0 \phi_1\rbx^{\H}\rbx}\nonumber\\
	&\times \sqbrc{\E_{\vbx'} \figbrc{\e^{-\xi_1 \norm{\rbz - \sqrt{\rho_1\beta_0}\rbx'}^2} \e^{\xi_1 \rbz^{\H}\rbz + \rho_1\beta_0 \psi_1 \rbx'^{\H}\rbx'}}}^u\Bigg\} \d \rbz.
\end{align}
Consequently, the generalized free energy in~\eqref{eqn:freeEnergyNew} reads
\begin{align}\label{eqn:freeEnergyMoments}
	\til \calF =&\;\!\! \int\! p_{\ir,\ii}\!(z;\eta_1) \frac{q_{\jr,\ji}\!(z;\xi_1)}{q_{0,0}\!(z;\xi_1)}\! \sqbrc{\frac{q_{1,0}\!(z;\xi_1)}{q_{0,0}\!(z;\xi_1)}}^{\lr}\!\! \sqbrc{\frac{q_{0,1}\!(z;\xi_1)}{q_{0,0}\!(z;\xi_1)}}^{\li}\!\!\!\!\! \d z,
\end{align}
where $p_{\ir,\ii}(z;\eta_1) \triangleq \E_{\vbx}\figbrc{(\Re\{x\})^{\ir} (\Im\{x\})^{\ii} p(z|x;\eta_1)}$ and $q_{\jr,\ji}(z;\xi_1) \triangleq \E_{\vbx'}\figbrc{(\Re\{x'\})^{\jr} (\Im\{x'\})^{\ji} q(z|x';\xi_1)}$.

The above expression reduces to the joint moments of $(\vx,\vx',\mmse{\vx'}_q)$, which, by the Carleman theorem~\cite[p.~227]{feller1971introduction}, implies the convergence in distribution.


\bibliographystyle{IEEEtran}
\bibliography{refsItTrans2014}

\end{document}